\documentclass[3p,twocolumn]{elsarticle}

\usepackage{lipsum}
\usepackage{amsmath,amssymb,amsfonts,bbold}
\usepackage{braket}
\usepackage{graphicx,subfigure}
\usepackage{hyperref}
\usepackage{mathtools}
\usepackage{comment}
\usepackage[normalem]{ulem}
\usepackage{cancel}

\newcommand{\be}{\begin{equation}}
\newcommand{\ee}{\end{equation}}
\newcommand{\bse}{\begin{subequations}}
\newcommand{\ese}{\end{subequations}}
\newcommand{\ba}{\begin{eqnarray}}
\newcommand{\ea}{\end{eqnarray}}
\newcommand{\bea}{\begin{eqnarray}}
\newcommand{\eea}{\end{eqnarray}}

\usepackage{color}
\newcommand{\addtext}[1]{{#1}}
\usepackage[normalem]{ulem}  

\usepackage{tikz}
\usetikzlibrary{patterns}

\begin{document}

\begin{frontmatter}

\title{Entanglement inequalities, black holes and the architecture of typical states}

\author[1]{Radouane Gannouji}
\ead{radouane.gannouji@pucv.cl}

\author[1]{Ayan Mukhopadhyay}
\ead{ayan.mukhopadhyay@pucv.cl}

\author[1,2]{Nicol\'{a}s Pinochet\corref{cor1}}
\ead{nicolas.pinochet.v@gmail.com}
\cortext[cor1]{Corresponding author}

\address[1]{Instituto de F\'{\i}sica, Pontificia Universidad Cat\'{o}lica de Valpara\'{\i}so, Avenida Universidad 330, Valpara\'{\i}so, Chile}
\address[2]{Departamento de F\'{\i}sica, Universidad T\'{e}cnica Federico Santa Mar\'{\i}a, Casilla 110-V, Valpara\'{\i}so, Chile}

\begin{abstract}
{
Using the holographic realization of the Araki--Lieb (AL) inequality, we argue that the leading entanglement plateau constrains the organization of typical pure states in a fixed-$(E,J)$ sector of a large-$c$ two-dimensional holographic CFT. On the field-theory side, its equality structure motivates an effective factorization of the regulated micro-canonical code subspace into ultraviolet (UV), intermediate and infrared factors, with part of the intermediate sector providing a purifier of the ultraviolet one. In the bulk, entanglement-wedge reconstruction realizes the ultraviolet sector as an asymptotic corona, universal to simple probes with length scale smaller than $L_{\rm UV}$. Crucially, $L_{\rm UV}$ is determined only by the conserved charges. At finite $c$, universal perturbative dressing correlates these leading factors and lifts the AL plateau at order $c^0$. We formulate a generalized ETH ansatz under which $O(e^{-S_<})$ fluctuations, with $S<=\min(S_L,S_R)$, account for the $O(c^0)$ deficit of the AL plateau, while the state dependence of matrix elements of individual simple operators smeared over lengths shorter than $L_{\rm UV}$ remains exponentially small. This provides a holographic realization and rotating generalization of the Eigenstate Thermalization Hypothesis: for zero rotation it reduces to the standard $O(e^{-S/2})$. Independently, the separation between $L_{\rm UV}$ and the scale of quantum correlations is of order $e^{-6S_</c}$ at large $S_<$.
}
\end{abstract}


\end{frontmatter}

\paragraph{Introduction:} { Thermalization of isolated quantum systems is best explained by the quantum-statistical properties of typical states, most sharply through the eigenstate thermalization hypothesis (ETH) \cite{DAlessio:2015qtq,Deutsch:2018ulr}. In large-$c$ holographic conformal theories, generic evolution of an isolated quantum system is described semiclassically by black-hole formation in anti-de Sitter space from non-trivial initial conditions \cite{Aharony:1999ti,Balasubramanian:2011ur,Chesler:2013lia}. It is therefore natural to ask whether the universal geometry of a black hole can reveal the architecture and statistical properties of the underlying typical pure states.

This raises a sharper question: can the organization of the degrees of freedom in typical high-energy pure states be revealed by the semi-classical black hole geometry via holographic entanglement measures, and can the same organization explain the emergence of ETH? We argue that the holographic Araki--Lieb plateau \cite{Hubeny:2013gta} provides such a link. Its saturation identifies a charge-dependent length scale $L_{\rm UV}$ below which the state has a universal structure, while the microscopic information distinguishing typical states is carried by longer-wavelength degrees of freedom. Holographically, this universal sector is represented by an asymptotic region of the black hole which we call the corona. At finite $c$ the separation is no longer exact, and correlations between these sectors lift the Araki--Lieb plateau at order $c^0$ \cite{Faulkner:2013ana,Chen:2017ahf}. We formulate a generalized ETH and argue that it naturally accounts for this lifting: exponentially small microscopic correlations can collectively produce order-one corrections to the Araki-Lieb plateau, while matrix elements of individual light operators supported on scales below $L_{\rm UV}$ remain exponentially close to their micro-canonical values. This suppression reduces to the standard ETH result in the absence of rotation -- for the rotating case, the state-dependence of the diagonal matrix elements are suppressed by the smaller chiral sector. We show that the semiclassical rotating black hole geometry reveals the generalized ETH suppression via the difference between $L_{\rm UV}$ and $L_{\rm QC}$, the scale of quantum correlations.

Our results are closely tied to holographic information recovery. The Page curve and entanglement-wedge reconstruction rely on an effective semiclassical separation between exterior and black-hole degrees of freedom \cite{PeningtonQES,AEMM,Penington:2019kki,AlmheiriQES,Almheiri2020Islands,Kibe:2021gtw,Akers:2022qdl,Kibe:2023ixa}. Our construction gives this separation a statistical interpretation: the leading split which isolates the asymptotic region (corona) from the black hole is fixed by the conserved charges, perturbative bulk correlations dress it at order $c^0$, while simple corona observables retain only exponentially smaller microscopic state dependence.}

\paragraph{Entanglement inequalities and their saturation:} We begin with a review of entanglement inequalities, and necessary and sufficient conditions for their saturation. Consider a state $\rho^{AB}$ in a bipartite Hilbert space $\mathcal{H} = \mathcal{H}_A \otimes \mathcal{H}_B$ with $A$ and $B$ denoting subsystems. $S(AB)$ denotes the von Neumann entropy of $\rho^{AB}$ i.e. $S(AB) = - \mathrm{Tr}(\rho^{AB} \ln \rho^{AB})$. The reduced density matrix of $A$ is defined as $\rho^A = \mathrm{Tr}_B(\rho^{AB})$ with partial trace in $\mathcal{H}_B$, and  $S(A)$, the entanglement entropy of $A$ is the von Neumann entropy of $\rho(A)$. Similarly, we can define $\rho^B$ and $S(B)$. The Araki-Lieb inequality \cite{Araki:1970ba,PhysRevLett.30.434,Lieb:1973cp} states that 
\begin{equation}
    \vert S(A) - S(B)\vert \leq S(AB).
\end{equation}
for any state $\rho^{AB}$ of the bipartite system. The condition for saturation of the inequality is related to other inequalities involving $J(A\vert B)$ and the quantum discord $D(A\vert B)$, which are measures of classical and quantum correlations between the two subsystems in the pure/mixed state $\rho^{AB}$, respectively, that are invariant under local unitary operations \cite{Zurek:2011vew,Ollivier:2001fdq,Henderson:2001wrr,Bera:2017lmd}. Firstly,
\begin{equation}
    J(A\vert B) = \max\limits_{\{E_i^B\}}\left[S(A) - \sum_i p_i S(\rho_i^A) \right]
\end{equation}
where the maximization is performed over all possible positive operator valued measures $\{E_i^B\}$ in $\mathcal{H}_B$ corresponding to measurements performed on $B$ with $\rho_i^A = {\rm Tr}_B(E_i^B\rho^{AB})/p_i$ denoting the post measurement state corresponding to the $i$th outcome of the measurement which has the probability $p_i = {\rm Tr}(E_i^B\rho^{AB})$. The quantum mutual information of $\rho^{AB}$ is defined as
\begin{equation}
    I(A:B) = S(A) + S(B) - S(AB).
\end{equation}
It is symmetric in $A$ and $B$, and equals to $2S(A) = 2S(B)$ iff $\rho^{AB}$ is a pure state. The quantum discord $D(A\vert B)$ defined as
\begin{equation}
    D(A\vert B) = I(A:B) - J(A\vert B)
\end{equation}
is generically asymmetric. {For pure states, however, it is symmetric and satisfies $D(A \vert B)=D(B \vert A)=\frac{1}{2} I(A: B)$}. It satisfies the inequality
\begin{equation}
    0 \leq D(A\vert B) \leq S(B)
\end{equation}
for any bipartite state $\rho^{AB}$ \cite{Zhang:2011hzq,PhysRevA.82.052122,2011JPhA...44K5301X}. The inequality
\begin{equation}\label{Eq:DJI}
    D(A\vert B)+J(C\vert B)\leq S(B),
\end{equation}
is valid for any tripartite state $\rho^{ABC}$, and is saturated if $\rho^{ABC}$ is pure, i.e. if the subsystem $C$ purifies $\rho^{AB}$ \cite{Koashi:2003pgf,Xi:2012ptf}. 

The upper bound on quantum discord is directly linked to the saturation of the Araki-Lieb inequality: the equality $D(A\vert B) = S(B)$ holds if and only if $S(A)-S(B)=S(AB)$. Obviously, this is realized only if $B$, the subsystem on which measurements are performed, has entropy not larger than $A$. Both maximal discord and saturation of the Araki-Lieb inequality occur iff $\mathcal{H}_A$ factorizes as $\mathcal{H}_A = \mathcal{H}_M \otimes \mathcal{H}_N$ such that
\begin{equation}\label{Eq:SS}
    \rho^{AB} = \rho^M \otimes \ket{\psi}^{NB}\bra{\psi}
\end{equation}
where $\rho^M $ is a (not necessarily pure) state in $\mathcal{H}_M$ and $\ket{\psi}^{NB}\bra{\psi}$ is a \textit{pure state} in $\mathcal{H}_N\otimes \mathcal{H}_B$, i.e. $\rho^B$ is necessarily purified by a subsystem of $A$ \cite{Xi:2012ptf,Bera:2017lmd}.

We readily note that for the state \eqref{Eq:SS}, we obtain
\begin{align}\label{Eq:EIs}
   & D(B\vert A) = D(A\vert B) = S(B) = S(A^N) \nonumber\\&=J(B\vert A) = J(A\vert B) =\frac{1}{2}I(A:B)
\end{align}
where $S(A^N) = S(B)$ is the von Neumann entropy of ${\rm Tr}_B(\ket{\psi}^{NB}\bra{\psi})$ \cite{Xi:2012ptf}. Note that for the state \eqref{Eq:SS}, both the quantum discord and classical correlations are symmetric in $A$ and $B$.

For a state \eqref{Eq:SS} saturating the Araki-Lieb inequality, let $C$ be its purifier. 
Such a pure state admits a quantum Markov chain \cite{Hayden:2003ytf} structure,
\begin{equation}\label{Eq:QMCS}
    \rho^{ABC} = \ket{\tilde{\psi}}^{CM}\bra{\tilde{\psi}}\otimes\ket{\psi}^{NB}\bra{\psi},
\end{equation}
with $C$ purifying $\rho^M$. Using \eqref{Eq:EIs}, we obtain from saturation of \eqref{Eq:DJI} that
\begin{align}\label{Eq:Id-S-1}
    &J(C\vert B)=D(C\vert B)=0,\nonumber\\ 
    &J(C\vert A)=D(C\vert A)=S(AB) =S(C).
\end{align}
The state \eqref{Eq:QMCS} also saturates strong subadditivity as
\begin{equation}\label{Eq:Id-S-2}
    S(CA)+S(AB)=S(A)+S(ABC)=S(A).
\end{equation}

\paragraph{The BTZ black hole:} 
Consider a two-dimensional holographic CFT on a circle of radius $L$ with equal central charge $c$ for the left and right movers. $L_{n,+}$ ($L_{n,-}$) are the Virasoro generators for the left (right) movers, respectively. 
The thermal state with temperature $\beta^{-1}$ and chemical potential $\beta\Omega$ for the momentum is given by
\begin{equation}\label{Eq:CE}
    \begin{split}
        \rho & =  \frac{ e^{-\beta(H -\Omega P)} }{ {\rm Tr}\!\left( e^{-\beta (H-\Omega P)} \right) }\\
        & =\frac{ e^{-\frac{\beta_+}{L} \left(L_{0,+}- \frac{c}{24}\right)} e^{-\frac{\beta_-}{L} \left(L_{0,-}-\frac{c}{24}\right)} }{ {\rm Tr}\left(  e^{-\frac{\beta_+}{L} \left(L_{0,+}- \frac{c}{24}\right)} e^{-\frac{\beta_-}{L} \left(L_{0,-}-\frac{c}{24}\right)} \right) },
    \end{split}
\end{equation}
where the Hamiltonian and momentum are 
\begin{align*}
H&=\frac{1}{L}\Bigl(L_{0,+}+L_{0,-}-\frac{c}{12}\Bigr),\\[-1mm]
P&=\frac{1}{L}\Bigl(L_{0,+}-L_{0,-}\Bigr).
\end{align*}
and
\begin{equation}\label{Eq:beta-omega-1}
    \beta = \frac{\beta_+ + \beta_-}{2}, \qquad 
    \Omega = \frac{\beta_- - \beta_+}{\beta_-+\beta_+}
\end{equation}
give the temperatures $\beta_+^{-1}$ ($\beta_-^{-1}$) of the left (right) movers. This state is holographically dual to the Ba\~nados-Teitelboim-Zanelli (BTZ) black hole \cite{Banados:1992wn}
\begin{align}\label{Eq:BTZ}
    {\rm d}s^2 
    = -\frac{\ell^2}{r^2 r_+^2 r_-^2}\,(r_+^2 - r^2)(r_-^2 - r^2)\,{\rm d}t^2 \nonumber \\
    + \frac{\ell^2 r_+^2 r_-^2}{r^2 (r_+^2 - r^2)(r_-^2 - r^2)}\,{\rm d}r^2 \nonumber \\
    + \frac{\ell^2}{r^2}\!\left( L\,{\rm d}\phi - \frac{r^2}{r_+ r_-}\,{\rm d}t \right)^{\!2},
\end{align}
which is locally AdS$_3$ spacetime with radius $\ell$ and with the conformal boundary at $r=0$. For $r_+<r_-$, the outer and inner horizons are at $r_+$ and $r_-$, respectively. The 
CFT central charge is given by $c=3\ell/2G_N$ (with large $c$ implying large N), where $G_N$ is the three-dimensional Newton constant. Let
\begin{equation}\label{Eq:mupm}
    \mu_{\pm}=\frac{1}{2}\left(\frac{1}{r_+}\pm\frac{1}{r_-}\right),
\end{equation}
and we set $\ell=1$ henceforth. With holographic renormalization \cite{Henningson:1998gx,Balasubramanian:1999re} we can extract the known non-vanishing components of the thermal stress tensor of the dual CFT with $\beta_\pm=\pi/\mu_\pm$. The thermal energy and momentum 
\begin{align}\label{Eq:EJ}
E&=\langle H\rangle=\frac{cL}{6}(\mu_+^2+\mu_-^2),\nonumber\\[-1mm]
J&=\langle P\rangle=\frac{cL}{6}(\mu_+^2-\mu_-^2).
\end{align}
correspond to the mass and momentum of the BTZ black hole whose horizon has Hawking temperature $\beta$ and rotation $\Omega$ 
\begin{equation}\label{Eq:beta-omega-2}
    \beta=\frac{\pi(\mu_++\mu_-)}{{2\mu_+\mu_-}},\qquad 
    \Omega=\frac{\mu_+-\mu_-}{\mu_++\mu_-}
\end{equation}
matching \eqref{Eq:beta-omega-1} also with $\beta_\pm=\pi/\mu_\pm$. The Bekenstein-Hawking entropy,
\begin{equation}\label{Eq:S}
    S_{BH}=S_L+S_R,\quad 
    S_L=\frac{c}{3}\pi L\mu_+,\quad 
    S_R=\frac{c}{3}\pi L\mu_-,
\end{equation}
of the BTZ black hole reproduces the quantum statistical entropy in the CFT in the large $c$ limit \cite{Cardy:1986ie,Strominger:1997eq}. For later convenience, we define
\begin{equation}\label{Eq:Sless}
    S_< = \min(S_L,S_R),
\end{equation}
so that $S_< = S_{BH}/2$ for $\Omega=0$ (no rotation). \addtext{For a rotating grand-canonical ensemble the BTZ saddle dominates in the thermal ensemble when
$\sqrt{\beta_+\beta_-}<2\pi L$; this reduces to $\beta<2\pi L$ at zero rotation} \cite{Hubeny:2013gta,Anous:2018hjh}. 

Results of numerical relativity \cite{Pretorius:2000yu,Choptuik:2004ha,Pandya:2020ejc,Bourg:2021vpv} and analytic solutions \cite{Danielsson:1999fa,Ross:1992ba,Matschull:1998rv,Lindgren:2015fum} suggest that BTZ black holes are formed as a result of gravitational collapse for generic initial conditions, at least when the momentum is small compared to the energy. The holographic duality implies that black hole formation mirrors the approach to typicality during unitary time evolution of an isolated system and that the BTZ black hole can quantify the statistical features of typical states.

\begin{figure}
\centering    

\tikzset{every picture/.style={line width=0.75pt}} 

\begin{tikzpicture}
[x=0.75pt,y=0.75pt,yscale=-1,xscale=1]
\begin{scope}[scale=0.55]
\draw  [draw opacity=0][fill={rgb, 255:red, 97; green, 180; blue, 15 }  ,fill opacity=0.06 ,even odd rule] (145.45,158.91) .. controls (145.45,106.7) and (187.78,64.37) .. (240,64.37) .. controls (292.21,64.37) and (334.54,106.7) .. (334.54,158.91) .. controls (334.54,211.13) and (292.21,253.45) .. (240,253.45) .. controls (187.78,253.45) and (145.45,211.13) .. (145.45,158.91)(93.09,158.91) .. controls (93.09,77.77) and (158.86,12) .. (240,12) .. controls (321.14,12) and (386.91,77.77) .. (386.91,158.91) .. controls (386.91,240.05) and (321.14,305.82) .. (240,305.82) .. controls (158.86,305.82) and (93.09,240.05) .. (93.09,158.91) ;
\draw  [color={rgb, 255:red, 0; green, 0; blue, 0 }  ,draw opacity=1 ][line width=1.5]  (93.09,158.91) .. controls (93.09,77.77) and (158.86,12) .. (240,12) .. controls (321.14,12) and (386.91,77.77) .. (386.91,158.91) .. controls (386.91,240.05) and (321.14,305.82) .. (240,305.82) .. controls (158.86,305.82) and (93.09,240.05) .. (93.09,158.91) -- cycle ;
\draw  [color={rgb, 255:red, 189; green, 16; blue, 224 }  ,draw opacity=1 ][fill={rgb, 255:red, 0; green, 0; blue, 0 }  ,fill opacity=0.59 ][dash pattern={on 3pt off 1.5pt}][line width=1.5]  (196.53,158.91) .. controls (196.53,134.9) and (215.99,115.44) .. (240,115.44) .. controls (264.01,115.44) and (283.47,134.9) .. (283.47,158.91) .. controls (283.47,182.92) and (264.01,202.38) .. (240,202.38) .. controls (215.99,202.38) and (196.53,182.92) .. (196.53,158.91) -- cycle ;
\draw  [dash pattern={on 0.75pt off 1.5pt}][line width=0.75]  (145.43,158.91) .. controls (145.43,106.69) and (187.77,64.35) .. (240,64.35) .. controls (292.22,64.35) and (334.56,106.69) .. (334.56,158.91) .. controls (334.56,211.14) and (292.22,253.48) .. (240,253.48) .. controls (187.77,253.48) and (145.43,211.14) .. (145.43,158.91) -- cycle ;
\draw [color={rgb, 255:red, 60; green, 190; blue, 31 }  ,draw opacity=1 ][line width=1.5]    (151.34,275.34) .. controls (155.36,270.52) and (159.15,263.69) .. (163.09,256.09) .. controls (167.03,248.49) and (171.67,236.01) .. (173.68,227.75) .. controls (175.7,219.49) and (175.75,218.51) .. (177.02,209.66) .. controls (178.3,200.81) and (178.8,188.75) .. (179.67,178.17) .. controls (180.54,167.6) and (182.99,156.36) .. (185.05,150.84) .. controls (187.11,145.33) and (187.82,143.05) .. (192.52,135.55) .. controls (197.23,128.05) and (203.66,122.95) .. (207.32,120) .. controls (210.98,117.05) and (216.24,114.01) .. (222.95,111.84) .. controls (229.67,109.67) and (234.01,109.09) .. (240.34,109.09) .. controls (246.67,109.09) and (252.83,110.59) .. (257.43,111.84) .. controls (262.03,113.09) and (266.17,115.42) .. (270.68,118.34) .. controls (275.19,121.26) and (279.84,125.75) .. (282.59,128.84) .. controls (285.34,131.93) and (290.09,138.8) .. (293.09,145.82) .. controls (296.09,152.84) and (298.43,162.24) .. (299.84,172.16) .. controls (301.25,182.08) and (301.34,193.39) .. (302.33,202.17) .. controls (303.33,210.96) and (304.59,221.18) .. (306.59,228.3) .. controls (308.59,235.41) and (311.34,242.8) .. (314.33,250.51) .. controls (317.33,258.22) and (322.52,266.47) .. (328.84,275.34) ;
\draw [color={rgb, 255:red, 208; green, 2; blue, 27 }  ,draw opacity=1 ][line width=1.5]  [dash pattern={on 3pt off 1.5pt}]  (151.34,275.34) .. controls (154.77,271.64) and (159.27,266.89) .. (165.02,263.89) .. controls (170.77,260.89) and (180.41,257.64) .. (184.34,256.64) .. controls (188.27,255.64) and (200.12,254.23) .. (209.73,253.64) .. controls (219.34,253.04) and (229.77,253.44) .. (240,253.48) .. controls (250.22,253.51) and (254.85,253.44) .. (262.57,253.64) .. controls (270.29,253.83) and (277.59,253.64) .. (285.43,254.64) .. controls (293.27,255.64) and (303.43,258.39) .. (312.09,262.39) .. controls (320.75,266.39) and (323.25,269.64) .. (328.84,275.34) ;
\draw    (402.82,151.29) -- (303.67,186.42) ;
\draw [shift={(301.79,187.08)}, rotate = 340.49] [color={rgb, 255:red, 0; green, 0; blue, 0 }  ][line width=0.75]    (10.93,-3.29) .. controls (6.95,-1.4) and (3.31,-0.3) .. (0,0) .. controls (3.31,0.3) and (6.95,1.4) .. (10.93,3.29)   ;
\draw    (403.49,205.96) -- (264.46,253) ;
\draw [shift={(262.57,253.64)}, rotate = 341.31] [color={rgb, 255:red, 0; green, 0; blue, 0 }  ][line width=0.75]    (10.93,-3.29) .. controls (6.95,-1.4) and (3.31,-0.3) .. (0,0) .. controls (3.31,0.3) and (6.95,1.4) .. (10.93,3.29)   ;
\draw    (405,100.55) -- (332.01,125.76) ;
\draw [shift={(330.12,126.42)}, rotate = 340.94] [color={rgb, 255:red, 0; green, 0; blue, 0 }  ][line width=0.75]    (10.93,-3.29) .. controls (6.95,-1.4) and (3.31,-0.3) .. (0,0) .. controls (3.31,0.3) and (6.95,1.4) .. (10.93,3.29)   ;
\draw [color={rgb, 255:red, 208; green, 2; blue, 27 }  ,draw opacity=1 ][line width=1.5]    (150.23,275.25) .. controls (155.48,279.19) and (161.93,283.68) .. (167.98,287) .. controls (174.03,290.32) and (183.32,294.53) .. (188.52,296.5) .. controls (193.72,298.47) and (202.05,301.32) .. (210.77,303) .. controls (219.49,304.68) and (234,306.14) .. (240.27,306) .. controls (246.54,305.86) and (254.42,305.45) .. (263.27,304) .. controls (272.13,302.55) and (286.11,298.9) .. (292.09,296.5) .. controls (298.08,294.1) and (301.09,292.75) .. (308.27,289) .. controls (315.45,285.25) and (322.66,280.37) .. (329.84,275.5) ;
\draw    (79.08,119.33) -- (193.36,158.68) ;
\draw [shift={(195.25,159.33)}, rotate = 199] [color={rgb, 255:red, 0; green, 0; blue, 0 }  ][line width=0.75]    (10.93,-3.29) .. controls (6.95,-1.4) and (3.31,-0.3) .. (0,0) .. controls (3.31,0.3) and (6.95,1.4) .. (10.93,3.29)   ;

\draw (299.73,48.03) node [anchor=north west][inner sep=0.75pt]    {$\mathfrak{C}$};
\draw (272.23,90.53) node [anchor=north west][inner sep=0.75pt]    {$\mathfrak{B}$};
\draw (406,140.95) node [anchor=north west][inner sep=0.75pt]    {$\mathcal{C}_{1}$};
\draw (406,192.95) node [anchor=north west][inner sep=0.75pt]    {$\mathcal{C}_{2}$};
\draw (406,92.95) node [anchor=north west][inner sep=0.75pt]    {$\Sigma $};
\draw (232,145.4) node [anchor=north west][inner sep=0.75pt]  [color={rgb, 255:red, 255; green, 255; blue, 255 }  ,opacity=1 ]  {$\mathfrak{I}$};
\draw (57.67,106.73) node [anchor=north west][inner sep=0.75pt]    {$\mathcal{H}$};

\end{scope}
\end{tikzpicture}

    \caption{The bulk geodesics ending on the boundary interval $R$ (red) are $\mathcal{C}_1$ (green) and $\mathcal{C}_2$ (dashed red). $\mathcal{C}_1\cup \mathcal{H}$ and $\mathcal{C}_2$ are homologous to $R$ with $\mathcal{H}$ the outer horizon (dashed magenta). $\mathcal{C}_1$ and $\mathcal{C}_2\cup \mathcal{H}$ are homologous to $\overline{R}$ (black), the complement of $R$. Here the BTZ black hole shown with compactified radial coordinate $r\to \cot\rho$ has $\mu_+=\mu_-=0.25$ (AdS radius $\ell =1$).}
    \label{fig:BiP}
\end{figure}
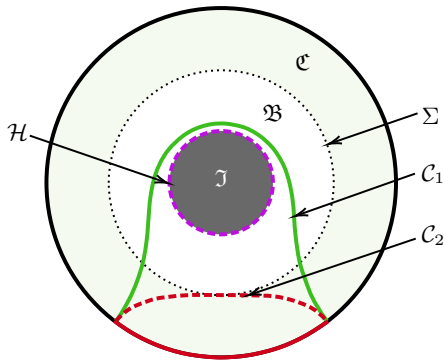

\paragraph{Holographic entanglement, the corona and the buffer:} 
In a large N holographic CFT, the entanglement entropy of a boundary subregion $R$ is given by the Hubeny-Rangamani-Ryu-Takayanagi (HRRT) prescription, which equates $S(R)$ to $1/(4G_N)$ times the area of co-dimension two bulk extremal surface(s) of least area and homologous to $R$ \cite{Rt,HRT}. This set of extremal surface(s) may have disconnected components but one must be anchored to $\partial R$. The entanglement wedge $E_W(R)$ is the bulk domain of dependence of a spatial region bounded by $R$ and its HRRT surface(s). Bulk operators in $E_W(R)$ can be reconstructed from CFT operators in $R$, and this is called \textit{entanglement wedge reconstruction} \cite{harlow2018tasi,Jafferis:2015del,Dong:2016eik,Jahn:2021uqr,Chen:2021lnq,Kibe:2021gtw}.

For the BTZ geometry \eqref{Eq:BTZ} whose boundary is a circle of length $2\pi L$, consider an interval $R$ of length $l<2\pi L$. The homology condition admits two competing configurations of extremal co-dimension two surfaces (curves): the geodesic $\mathcal{C}_2$, and a disconnected set consisting of the geodesic $\mathcal{C}_1$ together with the outer horizon $\mathcal{H}$ as shown in Fig.~\ref{fig:BiP}. The HRRT prescription yields
\begin{align}
    &S(R) =\frac{1}{4G_N} {\rm min} \left[ \mathbf{L}(\mathcal{C}_2),  \mathbf{L}(\mathcal{C}_1\cup \mathcal{H})\right],\nonumber\\
    &S(\overline{R}) =\frac{1}{4G_N} {\rm min} \left[  \mathbf{L}(\mathcal{C}_1),  \mathbf{L}(\mathcal{C}_2\cup \mathcal{H})\right],
\end{align}
where $ \mathbf{L}(C)$ denotes the length of a curve $C$ and $\overline{R}$ is the complement of $R$. Let us define 
\begin{equation}\label{Eq:R}
    \mathcal{P} = \frac{\vert S(R) - S(\overline{R}) \vert}{S(R\cup \overline{R})} = \frac{\vert S(R) - S(\overline{R}) \vert}{S_{BH}}
\end{equation}
with $S(R\cup\overline{R})=S_{BH}= \mathbf{L}(\mathcal{H})/(4G_N)$, the Bekenstein-Hawking entropy given by \eqref{Eq:S}.

Generically, for any BTZ geometry, there are three cases as shown in the Table \ref{tab:MinS} \cite{Hubeny:2013gta}.
\begin{table*}[t]
    \centering
    \begin{tabular}{cccc}
        \hline
         $l =  \mathbf{L}(R)$ & $S(R)$ &  $S(\overline{R})$  & $\mathcal{P}$\\
        \hline
         $0< l< 2\pi L(1-\alpha^*)$ & $\frac{ \mathbf{L}(\mathcal{C}_2)}{4 G_N}$
         & $\frac{ \mathbf{L}(\mathcal{C}_2)}{4G_N} + S_{BH}$ & 1\\
         $ 2\pi L(1-\alpha^*)\leq l < 2\pi L\alpha^* $ & $\frac{ \mathbf{L}(\mathcal{C}_2)}{4 G_N}$
         & $\frac{ \mathbf{L}(\mathcal{C}_1)}{4G_N}$ &  $< 1$ \\
         $2\pi L\alpha^* \leq l \leq 2\pi L$ & $\frac{ \mathbf{L}(\mathcal{C}_1)}{4 G_N} + S_{BH}$
         & $\frac{ \mathbf{L}(\mathcal{C}_1)}{4G_N}$ & 1\\
        \hline
    \end{tabular}
    \caption{Three cases of holographic entanglement in a generic BTZ geometry.}
    \label{tab:MinS}
\end{table*}
Note that $\mathcal{P}$ is symmetric under exchanging $R$ and $\overline{R}$, and so $\mathcal{P}=0$ for $l=\pi L$ ($\alpha^*\geq 0.5$ with equality only for $\mu_\pm =0$); note that the BTZ-dominance condition $\sqrt{\beta_+\beta_-}<2\pi L$ is assumed as discussed above. In the large-$c$ limit, we can thus define
\begin{align}
   &\alpha_{\rm UV} =1- \alpha^* \nonumber
   \\&= {\rm max}\left(\frac{\mathbf{L}(R)}{2\pi L}\,\bigg\vert\, S(\overline{R}) = S(R) + S(R\cup \overline{R})\right).
\end{align}
Thus the microscopic scale $L_{\rm UV}$ defined as
\begin{equation}
    L_{\rm UV} := 2\pi \alpha_{\rm UV} L, 
\end{equation}
is the \textit{largest} boundary interval whose entanglement wedge does \textit{not} include the outer horizon, and for which the Araki-Lieb inequality is \addtext{saturated by the leading HRRT entropy} ($\mathcal{P}=1$) in the large-$c$ limit. For any BTZ black hole, $\alpha_{\rm UV}$ (and thus $L_{\rm UV}$) can be obtained as a function of $\mu_+L$ and $\mu_-L$ by solving
\begin{align}\label{Eq:LUVEq}
&\ln \sinh(2\pi\mu_+L\alpha_{\rm UV})
 +\ln \sinh(2\pi\mu_-L\alpha_{\rm UV}) \nonumber\\
&\quad +2\pi L(\mu_++\mu_-) \nonumber\\
&=\ln \sinh\!\left(2\pi\mu_+L(1-\alpha_{\rm UV})\right) \nonumber\\
&\quad +\ln \sinh\!\left(2\pi\mu_-L(1-\alpha_{\rm UV})\right).
\end{align}
Here the logarithms on the left and right are respectively the renormalized geodesic-length contributions $\mathbf{L}(\mathcal C_2)-\ln(\mu_+\mu_-\epsilon^2)$ and $\mathbf{L}(\mathcal C_1)-\ln(\mu_+\mu_-\epsilon^2)$. 
The above equation can be obtained readily from explicit computations of the renormalized lengths of the geodesics $\mathcal{C}_1$ and $\mathcal{C}_2$ via uniformization maps \cite{HRT,Kibe:2024icu}. \addtext{Throughout this Letter, $L_{\rm UV}$ denotes this leading large-$c$, strong-coupling HRRT scale unless stated otherwise.}

We define the corona $\mathfrak{C}$ as the asymptotic region of the BTZ black hole that contains \textit{all} entanglement wedges corresponding to boundary subregions $R$ with length $l \leq L_{\rm UV}$. The buffer $\mathfrak{B}$ is the region between the outer horizon $\mathcal{H}$ and $\Sigma$, the inner boundary of the corona and the envelope of all entanglement wedges corresponding to subregions $R$ with length $l = L_{\rm UV}$ (see Fig. \ref{fig:BiP}).

\addtext{The HRRT entanglement plateau is exact only at leading order in the large-$c$ expansion. Bulk generalized-entropy corrections generically lift it already at order $c^0$ \cite{Faulkner:2013ana,Chen:2017ahf} so that for an interval $R$ inside the AL saturation plateau in the large-$c$ limit, the deficit is}
\begin{equation}
\addtext{\delta_{\rm AL}^{\rm leading}(R):=S_{BH}-\left[S(\overline R)-S(R)\right]=O(c^0).}
\label{Eq:AL-deficit-correction}
\end{equation}
\addtext{where $S_{BH}$ is the leading Bekenstein--Hawking entropy. In the bulk, the leading lifting is represented holographically by the mutual information of bulk quantum fields}
{
\begin{equation}
\delta_{\rm AL}^{\rm leading}(R)=I_{\rm bulk}\!\left(E_W(R):\mathfrak I\right)=O(c^0),
\label{Eq:AL-reference-identity}
\end{equation}
}
\addtext{where $\mathfrak I$ denotes the infrared/interior bulk degrees of freedom, $E_W(R)$ is the bulk entanglement wedge corresponding to the boundary interval $R$ \cite{Faulkner:2013ana,Chen:2017ahf}. Note that the bulk generalized-entropy receives both perturbative quantum and stringy corrections.}

\paragraph{An intuitive argument from Wilsonian RG:} {We now turn from the entanglement plateau to the organization of a typical pure state in the fixed-$(E,J)$ shell $\mathcal H_{E,J}$ in the large-$c$ limit. It is useful to begin with a heuristic picture based on the geometric realization of Wilsonian renormalization group (RG) flow in holography in the large-$c$ limit where the bulk radial coordinate can be identified with the energy scale of the dual theory \cite{Heemskerk:2010hk,Bredberg:2010ky,Faulkner:2010jy,Kuperstein:2011fn,Kuperstein:2013hqa,Behr:2015yna,Behr:2015aat,Mukhopadhyay:2016fre,Dharanipragada:2023mkc}. Based on this, we can argue that the degrees of freedom with characteristic wavelengths below $L_{\rm UV}$ are represented by the corona $\mathfrak C$; intermediate modes between $L_{\rm UV}$ and an infrared scale $L_{\rm IR}$  are associated with the buffer $\mathfrak B$; and sufficiently infrared degrees of freedom with characteristic wavelengths larger than $L_{\rm IR}$ are associated with the black-hole interior $\mathfrak I$. Phenomenologically, 
\begin{equation}
{
L_{\rm IR}:=\sqrt{\beta_+\beta_-}=\frac{\pi}{\sqrt{\mu_+\mu_-}},
}
\label{Eq:LIR}
\end{equation}
as $L_{\rm IR} < 2\pi L$ is needed so that the BTZ black hole saddle dominates in the gravitational path integral (see \cite{Anous:2018hjh}) at a fixed temperature. It is to be noted that both $L_{\rm UV}$ and $L_{\rm IR}$ are determined by the conserved charges and do not depend on the specific state in the micro-canonical shell. In what follows, we will formulate our arguments precisely without the need to invoke Wilsonian RG and $L_{\rm IR}$. We will proceed by making two sharp assumptions.

\textit{First Assumption:} There exists an effective regulated bipartition
\begin{equation}
\mathcal H_{E,J}=\mathcal H_{\overline{\mathfrak I}}\otimes\mathcal H_{\mathfrak I}
\end{equation}
such that, for a typical $\ket\psi\in\mathcal H_{E,J}$,
\begin{equation}\label{Eq:marginal}
\rho^{\overline{\mathfrak I}}_\psi={\rm Tr}_{\mathfrak I}(\ket\psi\bra\psi)
\end{equation}
is \textit{represented semiclassically by the geometry exterior to the outer horizon and has universal leading order entanglement spectrum fixed by $(E,J)$ in the large-$c$ limit}. The factor $\mathcal H_{\mathfrak I}$ is the CFT factor, which represents the infrared/interior degrees of freedom and purifies the exterior marginal within this regulated code-subspace description. The universal entanglement spectrum of $\rho^{\overline{\mathfrak I}}$ reproduces the black hole entropy at the leading order. As a consequence of this assumption, even the Renyi entropies of $\rho^{\overline{\mathfrak I}}_\psi$ for any typical $\ket{\psi}\in\mathcal H_{E,J}$ should be reproduced by the BTZ black hole geometry in a manner similar to how the semi-classical evapoating black hole geometry reproduces the entanglement spectrum of Hawking radiation \cite{PeningtonQES,AEMM,Penington:2019kki,AlmheiriQES,Almheiri2020Islands,Kibe:2021gtw,Akers:2022qdl,Kibe:2023ixa}.

The Jafferis-Lewkowycz-Maldacena-Suh (JLMS) relation and entanglement-wedge reconstruction identify the algebra of a boundary interval $R$ with that of its bulk entanglement wedge $E_W(R)$ \cite{Jafferis:2015del,Dong:2016eik,harlow2018tasi,Jahn:2021uqr,Chen:2021lnq,Kibe:2021gtw}. The union of all entanglement wedges corresponding to boundary intervals $R$ with length $l(R)\leq L_{\rm UV}$ defines $\mathcal H_{\mathfrak C}$, and via the JLMS the corresponding code-space factor in the dual CFT. Bulk locality then gives the effective regulated code-subspace factorization
\begin{equation}
\mathcal H_{\overline{\mathfrak I}}\simeq
\mathcal H_{\mathfrak C}\otimes\mathcal H_{\mathfrak B},
\end{equation}
with $H_{\mathfrak B}$ associated to the codespace for the buffer, and therefore
\begin{equation}
\mathcal H_{E,J}\simeq
\mathcal H_{\mathfrak C}\otimes\mathcal H_{\mathfrak B}\otimes\mathcal H_{\mathfrak I}.
\end{equation}
Furthermore, we obtain
\begin{equation}\label{Eq:marginal-1}
    \rho^{\overline{\mathfrak I}}_\psi =\rho^{\mathfrak{BC}}_\psi.
\end{equation}

Consider a boundary interval $R$ of length $l(R)\leq L_{\rm UV}$ . As stated above, $\mathcal{H}_{E_W(R)}$ in the bulk semiclassical effective theory can be identified with $\mathcal{H}_R$ by virtue of JLMS relation. Crucially, \addtext{at the leading order in the large-$c$ expansion}, the AL inequality is saturated, and $\overline{E_W}(R)$, the complement of $E_W(R)$ in the \textit{geometry exterior to the outer black hole horizon}, is then the entanglement wedge of $\overline{R}$ (as the extremal surface associated with $\overline{R}$ is the union of the extremal surface corresponding to $R$ and the outer horizon when $l(R)\leq L_{\rm UV}$). So, \addtext{at this leading order}, $\mathcal{H}_{\overline{E_W}(R)}$ can be identified with $\mathcal{H}_{\overline{R}}$. Note that the full geometry exterior to the outer horizon is the union of $E_W(R)$ and $\overline{E_W}(R)$ and is also the union of the buffer and the corona, and therefore
\begin{align}
    &\mathcal{H}_R \otimes  \mathcal{H}_{\overline{R}} \simeq \mathcal H_{\mathfrak C}\otimes\mathcal H_{\mathfrak B} \simeq \mathcal{H}_{\overline{\mathfrak I}}, \nonumber\\ &\mathcal H_{E,J}\simeq
\mathcal H_R\otimes\mathcal H_{\overline{R}}\otimes\mathcal H_{\mathfrak I}.
\end{align}
Consequently, we readily note from \eqref{Eq:marginal} and \eqref{Eq:marginal-1} that for any pure state $\ket\psi$ we obtain
\begin{align}
I(R:\mathfrak I) &= S(R) + S(\mathfrak I) - S(R\mathfrak I) \nonumber\\ &=
S(\mathfrak{BC})-[S(\overline R)-S(R)].
\label{Eq:purity-deficit}
\end{align}
using $ S(R\mathfrak I) = S(\overline R)$ and $S(\mathfrak I) =S(\mathfrak{BC})$, which are valid if $l(R) < L_{\rm UV}$. Since $S(R\overline{R})=S(\mathfrak{BC})=S_{BH}$ at the leading order owing to our first assumption, we note from \eqref{Eq:AL-deficit-correction} that the finite-$c$ version of the bulk AL-saturation deficit in the CFT is
\begin{equation}\label{Eq:AL-deficit-CFT}
    \delta_{\rm AL}(R) = I(R:\mathfrak{I}) 
\end{equation}
mirroring the bulk identity \eqref{Eq:AL-reference-identity} of \cite{Faulkner:2013ana,Chen:2017ahf}. Clearly, $I(R:\mathfrak{I}) = O(c^0)$.

\textit{Second Assumption:} At leading semiclassical order the buffer admits a further factorization
\begin{equation}
\mathcal H_{\mathfrak B}=\mathcal H_{\mathfrak L}\otimes\mathcal H_{\mathfrak K},
\end{equation}
such that $\mathcal H_{\mathfrak K}$ purifies the corona in a way fixed by $(E,J)$ and independent of the state in the micro-canonical shell. We do not associate separate spatial bulk regions with $\mathcal H_{\mathfrak L}$ and $\mathcal H_{\mathfrak K}$; this is an effective factorization of the regulated semiclassical code subspace.

\paragraph{The architecture of a typical state:} The two assumptions lead to the following leading architecture for an orthonormal family of microscopic states in the shell:
{
\begin{align}\label{Eq:psi-ach}
\ket{\psi_a^{(0)}}=
\left(\sum_i\sqrt{p_i}\,
\ket{\tilde\chi_{i,a}^{\mathfrak I}}\otimes
\ket{\chi_{i,a}^{\mathfrak L}}\right)
\otimes\ket\phi^{\mathfrak{KC}} .
\end{align}
}
Here $a$ labels a microscopic state in $ \mathcal{H}_{E,J}$; the two sets of $\chi$'s are orthogonal Schmidt bases; $p_i$ is a probability distribution fixed by the conserved charges at leading order; and the pure state $\ket\phi^{\mathfrak{KC}}$ is likewise independent of $a$ at this order. Thus the microscopic state dependence of the leading architecture resides in the long-wavelength Schmidt bases, whereas the corona together with its purifying factor is universal at fixed $(E,J)$.

The leading states in \eqref{Eq:psi-ach} are not the exact finite-$c$ states. Perturbative bulk quantum and stringy effects dress the complete regulated code subspace. We denote this schematically by
\begin{equation}
{
\ket{\psi_a}_{\rm EFT}\ \sim_{\rm sc}\
U_{\rm pert}(E,J)\ket{\psi_a^{(0)}} .
}
\label{Eq:psi-dressed}
\end{equation}
It is useful to explain why the dressing can be taken to be unitary. The leading semiclassical code subspace is an approximation to a subspace of the same underlying quantum theory, not a different Hilbert space with an independently chosen inner product. Perturbatively correcting the states must therefore preserve their Born inner products: if $\ket{\psi_a^{(0)}}$ and $\ket{\psi_b^{(0)}}$ are two leading states, the corresponding dressed states must satisfy, order by order,
\begin{equation}
{
{}_{\rm EFT}\!\langle\psi_a|\psi_b\rangle_{\rm EFT}
=\langle\psi_a^{(0)}|\psi_b^{(0)}\rangle
}
\label{Eq:inner-product-preservation}
\end{equation}
up to corrections beyond the perturbative order being retained. Thus the perturbative identification of the leading code subspace with the dressed one is an isometry. In a regulated finite-dimensional Hilbert space this isometry can be extended to a unitary transformation, so we may write
\begin{equation}
{
U_{\rm pert}=e^{iK_{\rm pert}},\qquad K_{\rm pert}^{\dagger}=K_{\rm pert}.
}
\label{Eq:Upert-unitary}
\end{equation}
For the degenerate fixed-$(E,J)$ shell the unitary mixes the microscopic basis states. Our genuinely nontrivial \textit{additional assumption} is not unitarity but \emph{universality}: the perturbative dressing is fixed by the common semiclassical BTZ background and the conserved charges rather than by the microscopic state label $a$. It need not preserve the tensor factors displayed in \eqref{Eq:psi-ach}; indeed, correlations generated by $U_{\rm pert}$ are precisely what can lift the classical AL plateau at order $c^0$. The symbol $\sim_{\rm sc}$ is therefore not a statement of closeness in Hilbert-space norm.

Tracing the leading state over $\mathcal H_{\mathfrak I}$ gives
{
\begin{align}\label{Eq:rho}
\rho^{\mathfrak{BC},(0)}_a=
\left(\sum_i p_i\ket{\chi_{i,a}^{\mathfrak L}}\bra{\chi_{i,a}^{\mathfrak L}}\right)
\otimes\ket\phi^{\mathfrak{KC}}\bra\phi .
\end{align}
}
At finite $c$ the reduced exterior state must be obtained by dressing the complete state and only then tracing the infrared factor. It is therefore convenient to define
\begin{equation}
{
\mathcal D_{\rm pert}^{E,J}[\varrho]
:=\operatorname{Tr}_{\mathfrak I}\!\left[
U_{\rm pert}(E,J)\varrho U_{\rm pert}^{\dagger}(E,J)
\right]
}
\label{Eq:Dpert-definition}
\end{equation}
for a density operator $\varrho$ on the full regulated $\mathfrak I\mathfrak B\mathfrak C$ code subspace. We then write
\begin{equation}
{
\rho^{\mathfrak{BC}}_a=
\mathcal D_{\rm pert}^{E,J}\!\left[
\ket{\psi_a^{(0)}}\bra{\psi_a^{(0)}}\right]
+\delta\rho^{\mathfrak{BC}}_{a,{\rm micro}} .
}
\label{Eq:rho-dressed}
\end{equation}
The first term contains the correlations generated by the universal perturbative dressing and can change collective entropies by $O(c^0)$. Equation~\eqref{Eq:rho-dressed} defines the remainder $\delta\rho^{\mathfrak{BC}}_{a,{\rm micro}}$ not captured by this common perturbative map. Since both terms whose difference defines it have unit trace, this remainder $\delta\rho^{\mathfrak{BC}}_{a,{\rm micro}}$ is Hermitian and traceless, but it need not be positive. More importantly, the notation should not be interpreted as saying that $\delta\rho^{\mathfrak{BC}}_{a,{\rm micro}}$ contains all of the microscopic state dependence, or that it is itself exponentially small in a strong norm: the first term in \eqref{Eq:rho-dressed} is also $a$-dependent because the same $U_{\rm pert}(E,J)$ acts on the leading state $\ket{\psi_a^{(0)}}$, in which the long wavelength bi-partition ($\mathfrak{IL}$) has Schmidt bases that are $a$-dependent as evident from \eqref{Eq:psi-ach}. We also define $\mathcal D_{\rm pert}$ on the complete state because $\mathfrak I$ is already entangled with $\mathfrak L$ at the leading order; in general, there is no autonomous channel acting only on the initial exterior marginal. 

The norm of $\delta\rho^{\mathfrak{BC}}_{a,{\rm micro}}$ can be also of $O(c^0)$. Nevertheless, the separation indicated in \eqref{Eq:rho-dressed} is useful for the generalized ETH relevant for light operators smeared over scales shorter than $L_{\rm UV}$, which will be stated later. As formulated below, generalized ETH constrains the microscopic state dependence of the restriction of the full exterior state $\rho^{\mathfrak{BC}}_a$ to $\mathcal A(R)$, the algebra of fixed light operators supported in an interval $R$ whose length is shorter than $L_{\rm UV}$. In particular, although $\delta\rho^{\mathfrak{BC}}_{a,{\rm micro}}$ can be of $O(c^0)$ in norm, the combined microscopic state dependence of the full exterior state $\rho^{\mathfrak{BC}}_a$ that can be resolved by any fixed operator in $\mathcal A(R)$ is only of $O(e^{-S<})$. The generalized ETH will provide a natural explanation for the $O(c^0)$ lifting of the AL plateau.

For the leading state \eqref{Eq:psi-ach}, $S(\mathfrak B)=S(\mathfrak C)+S(\mathfrak B\cup\mathfrak C)$, with $S(\mathfrak C)$ equal to the entropy of either member of the pure $\mathfrak K\mathfrak C$ factor. Therefore, $S(\mathfrak B\cup\mathfrak C)=S_{BH}=-\sum_i p_i\ln p_i + O(c^0)$ as desired. 
}

\paragraph{Explaining the \addtext{leading saturation} of Araki-Lieb inequality:} Consider a boundary interval $R$ of length $l(R)\leq L_{\rm UV}$. The saturation of the Araki-Lieb inequality at the leading order follows from the leading order architecture \eqref{Eq:psi-ach} of any typical state. We show this below.

We recall that since $l(R)\leq L_{\rm UV}$, the corresponding entanglement wedge of $R$ (which is $E_W(R)$) lies in the corona, and $\overline{E_W}(R)$, the complement of $E_W(R)$ in the geometry outside of the outer horizon in the BTZ black hole is the entanglement wedge of $\overline{R}$. Within the corona factor stated above, bulk locality lets us further factorize $\mathcal{H}_{\mathfrak{C}}$ as $\mathcal{H}_{\mathfrak{C}}=\mathcal{H}_{\overline{E_W^\mathfrak{C}}(R)}\otimes \mathcal{H}_{E_W(R)}$, with $E_W^\mathfrak{C}(R) = \overline{E_W}(R)\cap \mathfrak{C}$, the intersection of $\overline{E_W}(R)$ and the corona $\mathfrak{C}$, and $\mathcal{H}_{\overline{E_W^\mathfrak{C}}(R)}$ denoting the corresponding codespace in the dual CFT. Therefore, the \addtext{leading state \eqref{Eq:rho}} can be written as
{
\begin{align}\label{Eq:rho1}
    &\rho^{\mathfrak{BC},(0)}_{a} = \left(\sum_i p_i \ket{\chi_{i,a}^{\mathfrak{L}}}\bra{\chi_{i,a}^{\mathfrak{L}}}\right)\otimes \ket\phi^{\widetilde{\mathfrak{K}}E_W(R)}\bra\phi .
\end{align}
}
where  $\mathcal{H}_{\widetilde{\mathfrak{K}}}=\mathcal{H}_{\mathfrak{K}}\otimes\mathcal{H}_{\overline{E_W^\mathfrak{C}}(R)}$. Note that 
\begin{align*}
 &\mathcal{H}_{\mathfrak{L}}\otimes\mathcal{H}_{\widetilde{\mathfrak{K}}} = (\mathcal{H}_{\mathfrak{L}}\otimes \mathcal{H}_{\mathfrak{K}})\otimes\mathcal{H}_{\overline{E_W^\mathfrak{C}}(R)}\\& 
 = \mathcal{H}_{\mathfrak{B}}\otimes\mathcal{H}_{\overline{E_W^\mathfrak{C}}(R)}   = \mathcal{H}_{\overline{E_W}(R)},
\end{align*}
as $\overline{E_W}(R) = \mathfrak{B}\cup\overline{E_W^\mathfrak{C}}(R)$. The identifications of $\mathcal{H}_{\overline{E_W}(R)}$ with $\mathcal{H}_{\overline{R}}$ and $\mathcal{H}_{E_W(R)}$ with $\mathcal{H}_{R}$ imply that the \addtext{leading state \eqref{Eq:rho1} is} of the form \eqref{Eq:SS} that fulfills the necessary and sufficient condition for \addtext{leading saturation} of the Araki-Lieb inequality for the bi-partition $R$ and $\overline{R}$. Thus bulk locality allows us to rearrange the labels of the factorization of the leading order state \eqref{Eq:rho} obtained from our two assumptions and explain the leading order saturation of the Araki-Lieb inequality for all intervals $R$ shorter than $L_{\rm UV}$. At finite $c$, \eqref{Eq:AL-deficit-CFT} instead describes a perturbatively dressed state \eqref{Eq:rho-dressed} in which the Araki--Lieb plateau receives a correction of the order $O(c^0)$.

\paragraph{ETH as the statistical origin of the lifted Araki--Lieb plateau:} { We now return to the $O(c^0)$ lifting of the leading Araki--Lieb plateau. Our proposal is that generalized ETH provides its natural microscopic explanation. At leading order the architecture \eqref{Eq:psi-ach} factorizes the corona from the infrared degrees of freedom; the finite-$c$ dressing \eqref{Eq:psi-dressed} correlates these sectors. The central point is that an order-one collective mutual information does not require order-one microscopic deviations from micro-canonical behavior: exponentially many individually ETH-suppressed correlations can combine to give an $O(c^0)$ deficit of the Araki-Lieb plateau. We additionally formulate the generalized ETH suppression of the deviation of matrix elements of light operators that are smeared over intervals shorter than $L_{\rm UV}$ from their micro-canonical values if the perturbative dressing of the leading factorized architecture is state-independent as stated above. 

For a fixed microscopic state $a$ and an interval $R$ with $l(R)\leq L_{\rm UV}$, the regulated code-subspace identifications used in deriving \eqref{Eq:purity-deficit} give an effective tripartition $R\,\overline R\,\mathfrak I$, as shown above. We define
\begin{equation}
\rho_a^{R\mathfrak I}:=
\operatorname{Tr}_{\overline R}\!\left(\ket{\psi_a}\bra{\psi_a}\right),
\label{Eq:rho-RI-def}
\end{equation}
where $\ket{\psi_a}$ denotes the finite-$c$ state $\ket{\psi_a}_{\rm EFT}$ of \eqref{Eq:psi-dressed}. This is distinct from the exterior marginal $\rho_a^{\mathfrak{BC}}$ in \eqref{Eq:rho-dressed}: tracing $\mathfrak I$ gives the exterior state, whereas tracing $\overline R$ gives the state relevant for $I_a(R:\mathfrak I)$. Purity gives
\begin{equation}
I_a(R:\mathfrak I)
=S_a(\mathfrak{BC})-\bigl[S_a(\overline R)-S_a(R)\bigr].
\label{Eq:purity-deficit-a}
\end{equation}
We recall that \eqref{Eq:AL-deficit-CFT} identifies the finite-$c$ lifting of the plateau with this mutual information. The question we address is whether generalized ETH naturally reproduces its $O(c^0)$ scaling.

At leading order, \eqref{Eq:psi-ach} together with the AL decomposition \eqref{Eq:rho1} implies
\begin{equation}
\rho_a^{R\mathfrak I,(0)}
=\rho_R^{(0)}\otimes\rho_{\mathfrak I,a}^{(0)}
\equiv\sigma_a .
\label{Eq:BKM-product}
\end{equation}
Here $\rho_R^{(0)}$ is independent of the microscopic label $a$ at fixed $(E,J)$, whereas
\begin{equation}
\rho_{\mathfrak I,a}^{(0)}=
\sum_i p_i\ket{\widetilde\chi_{i,a}^{\mathfrak I}}\bra{\widetilde\chi_{i,a}^{\mathfrak I}}
\end{equation}
may depend on $a$ through its Schmidt basis while retaining the universal spectrum $\{p_i\}$. Equation~\eqref{Eq:BKM-product} is the information-theoretic form of leading AL saturation: an interval whose entanglement wedge lies in the corona is uncorrelated with the infrared factor at this order.

To relate the local information geometry at $\sigma_a$ to the actual finite-$c$ state, we fix the normalization of the tangent by choosing the straight-line interpolation
\begin{equation}
\rho_a(t)
:=(1-t)\sigma_a+t\rho_a^{R\mathfrak I},
\qquad 0\leq t\leq1 .
\label{Eq:BKM-family}
\end{equation}
This is a valid curve of density matrices and satisfies
\begin{equation}
X_a:=\frac{d\rho_a(t)}{dt}
=\rho_a^{R\mathfrak I}-\sigma_a,
\qquad \operatorname{Tr}X_a=0 .
\label{Eq:physical-tangent}
\end{equation}
(Note that $\operatorname{Tr}X_a=0$ is simply a consequence of $\operatorname{Tr}(\rho_a^{R\mathfrak I}(t)) =1$ for all $t$.) Thus $X_a$ is the displacement from the leading AL-saturating state towards the physical finite-$c$ state. The parameter $t$ is only a coordinate along this interpolation; in particular, it is not $1/c$, $1/\sqrt c$, or any other large-$c$ expansion parameter. Choosing the physical endpoint merely fixes the normalization of the tangent and does not amount to evaluating a quadratic Taylor approximation at $t=1$.

The marginal tangents are
\begin{align}
X_{R,a}&:=\operatorname{Tr}_{\mathfrak I}X_a
=\rho_{R,a}-\rho_R^{(0)},\nonumber\\[-1mm]
X_{\mathfrak I,a}&:=\operatorname{Tr}_{R}X_a
=\rho_{\mathfrak I,a}-\rho_{\mathfrak I,a}^{(0)}.
\end{align}
We separate the tangent that only changes the two marginals,
\begin{equation}
X_{a,{\rm prod}}
=X_{R,a}\otimes\rho_{\mathfrak I,a}^{(0)}
+\rho_R^{(0)}\otimes X_{\mathfrak I,a},
\end{equation}
from the genuinely correlating part
\begin{equation}
X_{a,{\rm conn}}:=X_a-X_{a,{\rm prod}}.
\label{Eq:BKM-connected}
\end{equation}
Its relation to the connected part of the physical density matrix is explicit. As shown in Appendix~\ref{Sec:BKMTheorem}, defining
\begin{equation}
C_a:=\rho_a^{R\mathfrak I}-\rho_{R,a}\otimes\rho_{\mathfrak I,a},
\end{equation}
one finds exactly that
\begin{equation}
C_a=X_{a,{\rm conn}}-X_{R,a}\otimes X_{\mathfrak I,a}.
\label{Eq:C-Xconn}
\end{equation}
Hence $X_{a,{\rm conn}}$ is the linearized connected correlation generated by the physical finite-$c$ displacement; the difference between $C_a$ and $X_{a,{\rm conn}}$ starts at quadratic order in that displacement.

As
\begin{equation}
\mathcal I_a(t) :=  I_{\rho_a(t)}(R:\mathfrak I)=D\!\left(\rho_a(t)\,\middle\|\,\rho_{R,a}(t)\otimes\rho_{\mathfrak I,a}(t)\right),
\label{Eq:BKM-relative}
\end{equation}
vanishes at $t=0$ (with $\rho_{R,a}(0)$ independent of the label $a$), the standard stationarity and Bogoliubov--Kubo--Mori Hessian properties of relative entropy \cite{Petz1996Monotone,LesniewskiRuskai1999}, proved in Appendix~\ref{Sec:BKMTheorem}, yields
\begin{align}
\dot{\mathcal I}_a(0)&=0,
\label{Eq:BKM-stationarity}\\[1mm]
\ddot{\mathcal I}_a(0)
&=\operatorname{Tr}\!\left[
X_{a,{\rm conn}}\,
\Omega_{\sigma_a}^{-1}(X_{a,{\rm conn}})\right]\geq0,
\label{Eq:BKM-Hessian}
\end{align}
where
\begin{equation}
\Omega_{\sigma_a}^{-1}(Y)
:=\int_0^\infty ds\,(\sigma_a+s)^{-1}Y(\sigma_a+s)^{-1}.
\label{Eq:BKM-omega}
\end{equation}
We will refer to the above results as the BKM theorem. In what follows, we will use the corresponding BKM inner product and norm,
\begin{align}
\langle Y,Z\rangle_{{\rm BKM},\sigma_a}
&:=\operatorname{Tr}\!\left[Y\,\Omega_{\sigma_a}^{-1}(Z)\right],\nonumber\\[-1mm]
\|Y\|_{{\rm BKM},\sigma_a}^{2}
&:=\langle Y,Y\rangle_{{\rm BKM},\sigma_a}.
\label{Eq:BKM-norm}
\end{align}
Because $\mathcal I_a(0)=0$ and $\dot{\mathcal I}_a(0)=0$, the second equation above fixes the quadratic coefficient in the expansion about the AL-saturating point $t=0$, so that
\begin{equation}
\mathcal I_a(t)
=\frac{t^2}{2}\,
\|X_{a,{\rm conn}}\|_{{\rm BKM},\sigma_a}^{2}+O(t^3).
\label{Eq:BKM-Taylor}
\end{equation}
This equation is the precise bridge between the BKM theorem and the statistical argument below. However, it is a local expansion around $t=0$; so we cannot readily use it to estimate $\mathcal I_a(1)$ which is our target as
\begin{equation}
\mathcal I_a(1)=I_a(R:\mathfrak I)=D\!\left(\rho_a^{R\mathfrak I}\,\middle\|\,\rho_{R,a}\otimes\rho_{\mathfrak I,a}\right).
\end{equation}

To proceed further, we expand the full connected tangent for each fixed $a$ in Hermitian BKM-orthonormal directions,
\begin{align}
X_{a,{\rm conn}}
&=\sum_{\alpha=1}^{N_{\rm eff}}x_{\alpha,a}T_{\alpha,a},\nonumber\\[-1mm]
\langle T_{\alpha,a},T_{\beta,a}\rangle_{{\rm BKM},\sigma_a}
&=\delta_{\alpha\beta}.
\label{Eq:rho-micro-expansion}
\end{align}
The coefficients $x_{\alpha,a}$ are real, and the index $a$ on the basis vectors only reminds us that the BKM metric is evaluated at $\sigma_a$.

For a rotating two-dimensional CFT, the chiral sectors have degeneracies $D_L\sim e^{S_L}$ and $D_R\sim e^{S_R}$. Assuming these correlations have covariance across both chiral sectors, they are bottlenecked by the smaller Schmidt support:
\[
D_<:=\min(D_L,D_R)=e^{S_<},\,\,
S_<:=\min(S_L,S_R),
\]
so the number of independent correlation directions scales parametrically as
\begin{equation}
N_{\rm eff}\sim D_<^2=e^{2S_<}.
\label{Eq:Neff-chiral}
\end{equation}
By generalized ETH, we mean that a typical BKM-normalized component $|x_{\alpha,a}|$ has the RMS size
\begin{equation}
|x_{\alpha,a}| \sim x_{\rm rms}\sim D_<^{-1}=N_{\rm eff}^{-1/2}=e^{-S_<}.
\label{Eq:ETH-scale-from-counting}
\end{equation}
For zero rotation, $S_L=S_R=S_{BH}/2$, and we recover the standard $e^{-S_{BH}/2}$ ETH suppression.

Let
\begin{equation}
\mathcal I_a(t)=\sum_{n\geq2}t^n I_a^{(n)}.
\label{Eq:Ia-series}
\end{equation}
Eq.~\eqref{Eq:BKM-Taylor} then immediately gives the quadratic coefficient along the physical interpolation,
\begin{align}
&I_a^{(2)}(R:\mathfrak I)
=\frac12\|X_{a,{\rm conn}}\|_{{\rm BKM},\sigma_a}^{2}=\frac12\sum_{\alpha=1}^{N_{\rm eff}}|x_{\alpha,a}|^2\nonumber\\[-1mm]
&
\sim \frac12N_{\rm eff}x_{\rm rms}^2\sim \frac12N_{\rm eff}N_{\rm eff}^{-1}
=O(1).
\label{Eq:Ia-quadratic-BKM}
\end{align}
Thus the first non-vanishing information-geometric response is already of order $c^0$: the ETH suppression of each component is compensated by the exponentially large number of independent directions.

This quadratic coefficient need not numerically dominate the physical value at $t=1$, since the total displacement can have order-one BKM norm. 
However, we can use the usual quasi-Gaussian/random-phase part of the ETH ansatz which states that the $x_{\alpha,a}$ have no exponentially coherent alignment. For an ideal Gaussian ensemble odd connected moments vanish, while at order $2m$ there are at most $m$ independent pairings, giving parametrically
\begin{equation}
I_a^{(2m)}\sim N_{\rm eff}^{m}x_{\rm rms}^{2m}=O(1),
\qquad m\geq1.
\label{Eq:Ia-higher-order-counting}
\end{equation}
Hence no fixed order grows exponentially with $S_<$. Provided the expansion \eqref{Eq:Ia-series} admits a regular order-one resummation at the physical endpoint $t=1$, we obtain
\begin{equation}
I_a(R:\mathfrak I)=\mathcal I_a(1)=O(c^0).
\label{Eq:Ia-statistical-estimate}
\end{equation}
This is the sense in which generalized ETH explains the lifted Araki--Lieb plateau: exponentially small typical components of the connected correlation collectively give an order-one mutual information, which by \eqref{Eq:purity-deficit-a} and \eqref{Eq:AL-deficit-CFT} is precisely the $O(c^0)$ failure of leading AL saturation.

The universality of $U_{\rm pert}(E,J)$ becomes essential when we pass from this collective statement to operator ETH. For a fixed light boundary operator $O_R\in\mathcal A(R)$ smeared over the interval $R$, the same dressing acts on every state in the fixed-$(E,J)$ shell, so
\begin{align}
&{}_{\rm EFT}\!\bra{\psi_b}O_R\ket{\psi_a}_{\rm EFT}
=\bra{\psi_b^{(0)}}O_R^{\rm pert}\ket{\psi_a^{(0)}},
\nonumber\\[-1mm]
&O_R^{\rm pert}:=U_{\rm pert}^{\dagger}(E,J)O_RU_{\rm pert}(E,J).
\label{Eq:Opert-common}
\end{align}
Thus the same dressed observable is tested on every state in the shell. This is precisely where the $a$-independence of $U_{\rm pert}$ is needed. If the dressing were state dependent, even the diagonal matrix element would involve a different operator $U_{{\rm pert},a}^{\dagger}O_RU_{{\rm pert},a}$ for each $a$, while an off-diagonal matrix element would contain $U_{{\rm pert},b}^{\dagger}O_RU_{{\rm pert},a}$. The microscopic label could then be hidden in the dressing itself, and the usual ETH statement for one fixed observable would no longer follow from the state architecture.

For any pair of labels $a,b$ denoting  states in the micro-canonical shell define the reduced transition operator
\[
\rho_R^{ab}:=\operatorname{Tr}_{\overline R\mathfrak I}
\bigl(\ket{\psi_a}\bra{\psi_b}\bigr).
\]
Equivalently, with
\[
\rho_{ab}^{\mathfrak{BC}}:=\operatorname{Tr}_{\mathfrak I}
\bigl(\ket{\psi_a}\bra{\psi_b}\bigr),
\qquad
\rho_{ab}^{R\mathfrak I}:=\operatorname{Tr}_{\overline R}
\bigl(\ket{\psi_a}\bra{\psi_b}\bigr),
\]
we have, because $\mathcal{H}_R \otimes  \mathcal{H}_{\overline{R}} \simeq \mathcal H_{\mathfrak C}\otimes\mathcal H_{\mathfrak B}$,
\[
\rho_R^{ab}
=\operatorname{Tr}_{\overline R}\rho_{ab}^{\mathfrak{BC}}
=\operatorname{Tr}_{\mathfrak I}\rho_{ab}^{R\mathfrak I},
\,\,\,\,
\operatorname{Tr}_R(O_R\rho_R^{ab})
=\langle\psi_b|O_R|\psi_a\rangle .
\]
For $a=b$, $\rho_R^{aa}=\rho_{R,a}$ is therefore simultaneously the $R$-marginal of the exterior state in \eqref{Eq:rho-dressed} and of the $R\mathfrak I$ state used in the BKM analysis. The two ETH statements discussed here probe different projections of the same finite-$c$ state: $X_{a,{\rm conn}}$ controls the collective $R$--$\mathfrak I$ mutual information, whereas operator ETH concerns the marginal on $R$. Indeed,
\[
\operatorname{Tr}_{\mathfrak I}X_{a,{\rm conn}}=0,
\qquad
X_{R,a}=\rho_{R,a}-\rho_R^{(0)}.
\]

For $l(R)<L_{\rm UV}$, the leading AL structure \eqref{Eq:BKM-product} makes $\rho_R^{(0)}$ independent of $a$. We define the micro-canonical reduced state
\begin{align*}
    \rho_R^{\rm mc}:=\frac{1}{\dim\mathcal H_{E,J}}
\sum_a\rho_R^{aa},\\
O_{\rm univ}(E,J):=\operatorname{Tr}_R(O_R\rho_R^{\rm mc}).
\end{align*}
In a rotating ensemble, the diagonal and off-diagonal matrix elements probe different statistical aspects of the same reduced state and need not have the same suppression. The diagonal part measures the microscopic state dependence of the corona marginal itself. Motivated by the smaller-chiral-sector bottleneck in the connected correlations discussed above, we take this scale to be controlled by $S_<$. Generic off-diagonal matrix elements, on the other hand, sample the full fixed-$(E,J)$ shell and are normalized by its total entropy. Accordingly, for every fixed normalized light operator $O_R\in\mathcal A(R)$ we formulate
\begin{align}
\operatorname{Tr}_R\!\left[
O_R\left(\rho_R^{aa}-\rho_R^{\rm mc}\right)
\right]
&=\langle\psi_a|O_R|\psi_a\rangle-O_{\rm univ}(E,J)\nonumber\\
&=O_{\rm RMS}(e^{-S_<}),\nonumber\\[-1mm]
\operatorname{Tr}_R\!\left(O_R\rho_R^{ab}\right)
&=\langle\psi_b|O_R|\psi_a\rangle\nonumber\\
&=O_{\rm RMS}(e^{-S_{BH}/2}),\qquad a\neq b,
\label{Eq:observable-ETH-corrected}
\end{align}
with $l(R)<L_{\rm UV}$. Since $S_{BH}=S_L+S_R$, the generic off-diagonal scale is $e^{-(S_L+S_R)/2}$.

It is easy to show why the normalization of the off-diagonal matrix elements is fixed by the complete fixed-$(E,J)$ shell. Writing $D=\dim\mathcal H_{E,J}\sim e^{S_{BH}}$ and $M_{ab}=\langle\psi_b|O_R|\psi_a\rangle$, one has
\begin{align*}
&\frac{1}{D}\sum_{a,b}|M_{ab}|^2
=\frac{1}{D}\operatorname{Tr}\!\left(P_{E,J}O_R^\dagger P_{E,J}O_RP_{E,J}\right)
\\
&\leq \langle O_R^\dagger O_R\rangle_{\rm mc}=O(c^0).
\end{align*}
Thus, for a generic light operator whose matrix is dense in the symmetry-resolved shell, the usual ETH typicality estimate gives an off-diagonal RMS of order $D^{-1/2}=e^{-S_{BH}/2}$. At zero rotation $S_L=S_R=S_{BH}/2$, so the diagonal and off-diagonal suppressions coincide with the familiar $e^{-S_{BH}/2}$ scale. Special chiral operators or conserved currents may obey additional selection rules and hence be sparse in the full shell, in which case the scaling of their non-vanishing matrix elements need not follow this generic dense-shell estimate.

As already hinted before, Equation~\eqref{Eq:observable-ETH-corrected} should not be interpreted in terms of the norm of the deviation $\delta\rho^{\mathfrak{BC}}_{a,{\rm micro}}$ from the perturbatively dressed leading order factorized form of the state $\rho^{\mathfrak{BC}}_a$ represented by the black hole geometry exterior to the outer horizon. It is an operational statement about the restriction of this exterior state to the algebra of fixed light corona observables. It is also fully compatible with the order-one collective information in \eqref{Eq:Ia-statistical-estimate}: the $e^{-S_<}$ scale in the BKM argument refers to BKM-normalized correlation directions, whereas generic off-diagonal matrix elements of a fixed light operator are normalized over the full micro-canonical shell. 

If the support of the operator extends beyond the AL-protected corona scale, the leading universality of $\rho_R^{(0)}$ no longer follows from the AL argument, and no corresponding diagonal suppression follows from this reasoning. Thus AL saturation identifies the class of observables for which the leading reduced state is universal and motivates the $e^{-S_<}$ diagonal generalized-ETH scale, while the $a$-independence of $U_{\rm pert}$ ensures that every state in the shell is compared with the same perturbatively dressed observable. Together with the $e^{-S_{BH}/2}$ of generic off-diagonal matrix elements, this gives the generalized ETH statement for fixed light operators smeared over length scales shorter than $L_{\rm UV}$, which is determined by the conserved charges.

\paragraph{Geometry and the statistics of ETH suppression:} Finally, the rotating BTZ black hole geometry provides an independent check of the same physical picture. In order to connect to ETH, we can consider two distinct geometric quantities in the rotating BTZ black hole geometry, each of which has a distinct physical meaning in terms of quantum information property of a typical state, such that they must agree in the thermodynamic limit. The proposed leading order factorized architecture \eqref{Eq:psi-ach} and its universal perturbative dressing \eqref{Eq:psi-dressed} of a typical state suggest that two such geometric quantities can be $L_{\rm UV}$ and $L_{\rm QC}$, the length scale of quantum correlations. As defined precisely in Appendix~\ref{Sec:Supp}, $L_{\rm QC}$ is the maximum length of the smaller of the two gaps between two disjoint intervals for which these two intervals have non-vanishing mutual information. Since only the UV pure state factor in \eqref{Eq:psi-ach} is state-independent and the perturbative unitary dressing in \eqref{Eq:psi-dressed} is also state-independent, we expect $L_{\rm QC}$ to approach $L_{\rm UV}$ after state-averaging, up to corrections controlled by the same smaller chiral sector that governs diagonal generalized ETH since the entanglement measures defining both scales are represented by diagonal expectation values of twist-operator products.

Therefore, we expect that at large $S_<$ we should have
\begin{equation}
\frac{L_{\rm UV}-L_{\rm QC}}{L_{\rm UV}}
\propto e^{-2\pi\mu_<L}
=e^{-6S_</c}.
\label{Eq:geometric-heuristic}
\end{equation}
Appendix~\ref{Sec:Supp} verifies this explicitly at large $S_<$:
\begin{align}
\frac{L_{\rm UV}-L_{\rm QC}}{L_{\rm UV}}
&\sim A(r)e^{-6S_</c},\nonumber\\[-1mm]
r&=\frac{\mu_<}{\mu_>}\in(0,1),
\label{Eq:rotating-geometric-asymptotic}
\end{align}
with $A(r)$ independent of the circle size, $\mu_<:=\min(\mu_+,\mu_-)$ and $\mu_>:=\max(\mu_+,\mu_-)$. At zero rotation $A(1)=1/\ln2$, while $A(1/2)\simeq0.574$. 

At extremality $S_<=0$, and our arguments for the ETH origin of the $O(c^0)$ correction to the Araki-Lieb plateau presented above suggest that we should not expect any generalized ETH near extremality as $N_{\rm eff}$ should be related to an effectively finite number of zero/soft modes of the scale invariant near-horizon geometry \cite{Mertens:2022irh,Poojary:2022meo,Kapec:2024zdj,Despontin:2026xzg}. Indeed, the correction to $S_{BH}$ is macroscopic at extremality, and we can expect a macroscopic $O(c)$ correction to the Araki-Lieb plateau also which should be interesting to verify explicitly in the future. We do find that at extremality 
\[
L_{\rm UV}-L_{\rm QC} 
\sim 0.33 \tilde{\mu}^{-1}
\]
where $\tilde{\mu}$ is the non-vanishing member of the set $\{\mu_-, \mu_+\}$. Therefore, $L_{\rm UV}-L_{\rm QC} $ suggests that we can expect that at extremality, the deviation of matrix elements of light operators smeared over scales smaller than $L_{\rm UV}$ from their values in the micro-canonical ensemble should be suppressed by inverse power of the entropy instead of the suppression implied by generalized ETH. In absence of the latter, there is no reason to expect that the correction to the entanglement plateau will be only $O(c^0)$.

\paragraph{Discussion:} In this letter, we have shown that the saturation of the Araki-Lieb plateau in holographic two-dimensional CFTs and its $O(c^0)$ correction in the large-$c$ limit can be explained from a leading order factorization of a typical state, its state-independent unitary dressing which does not preserve these factors and a generalization of ETH which includes rotating ensembles and reduces to the standard ETH in absence of rotation. Crucially, we have identified a length scale $L_{\rm UV}$ determined by the saturation of the Araki-Lieb inequality at the leading order, given only in terms of the conserved charges. The algebra of light operators smeared over intervals with lengths shorter than $L_{\rm UV}$ satisfies generalized ETH when the entropy of the smaller chiral sector is large. 

Our results have significant consequences for microstates of black holes far from extremality. The corona of the black hole formed by the union of entanglement wedges of intervals shorter than $L_{\rm UV}$ acquires a precise operational meaning. Simple probes whose support lies strictly inside $L_{\rm UV}$ at the boundary see the universally dressed corona, with only exponentially small residual dependence on the black-hole microstate. Collective quantities are different: correlations between the interior and corona can be $O(c^0)$ while remaining parametrically subleading to $S_{BH}$. Correspondingly, we can draw the following conclusions  for the collective correlations:
\begin{equation}
J(\mathfrak I\vert\mathfrak C),D(\mathfrak I\vert\mathfrak C)=O(c^0),
\end{equation}
and
\begin{equation}
J(\mathfrak I\vert\mathfrak B),D(\mathfrak I\vert\mathfrak B)=S_{BH}+O(c^0),
\end{equation}
from our earlier discussion, while fixed simple corona observables retain the much smaller $O(e^{-S_<})$ microscopic dependence implied by generalized ETH. Such a picture is compatible with microstate geometries and other realizations of quantum hair \cite{Mathur:2005zp,Chakraborty:2025jtj,Bena:2018bbd,Bena:2019azk,Bena:2024gmp,Bena:2025uyg,Emparan:2025ymx,Bena:2025hxt}.

Several questions remain open. It would be useful to understand whether state-dependent proposals for interior reconstruction \cite{Papadodimas:2013wnh,Papadodimas:2013jku,Hayden:2018khn,Akers:2022qdl} can be reformulated directly in terms of the infrared factor and its universally dressed exterior purifier, and how perturbative entanglement-wedge reconstruction \cite{Almheiri:2014lwa,Cotler:2017erl,Bahiru:2022oas,Bahiru:2023zlc} implements the $O(c^0)$ correlations discussed here. A dynamical version of the architecture in BTZ--Vaidya geometries would allow one to follow the corona and the microscopic suppression scale through a quench \cite{Kibe:2021qjy,Banerjee:2022dgv,Kibe:2024icu}. It would also be interesting to determine the analogue of the chiral bottleneck for rotating black holes in higher dimensions.

Furthermore, the identification of the asymptotic corona region (UV sector) that isolates the black hole can lead to applications of holography to realistic strongly interacting systems \cite{Hartnoll:2016apf,Baggioli:2019rrs}, especially via the semi-holographic approach \cite{Faulkner:2010tq,Iancu:2014ava,Mukhopadhyay:2013dqa,Banerjee:2017ozx,Kurkela:2018dku,Ecker:2018ucc,Doucot:2024hzq}, where the UV sector is replaced by perturbative degrees of freedom. It would also be valuable to understand how the architecture \eqref{Eq:psi-ach}, its universal perturbative dressing and the $e^{-S_<}$ microscopic fluctuations are represented in tensor-network constructions \cite{Geng:2025efs,Qasim:2025qqm,Balasubramanian:2025rcr}. These questions may clarify more generally how semiclassical bulk locality emerges statistically from the structure of typical quantum states.

\paragraph{Acknowledgments:}  We are especially grateful to Tanay Kibe as collaboration between him, AM and NP on deriving the results of this paper purely from CFT has informed the final form of this work. We also thank Iosif Bena, Rapha\"{e}l Dulac, Matthew Headrick, Alexander Jahn, Feng-Li Lin, Giuseppe Policastro and Shozab Qasim for very valuable discussions. RG, AM and NP acknowledge support from FONDECYT regular grant no. 1220965, FONDECYT regular grant no. 1240955 and ``Doctorado Nacional'' grant no. 21221414 of La Agencia Nacional de Investigaci\'{o}n y Desarrollo (ANID), Chile, respectively. AM gratefully acknowledges the hospitality of LPENS, where a substantial part of this work was carried out during his tenure as a CNRS invited professor.

\appendix

\section{Stationarity and the BKM Hessian of mutual information}\label{Sec:BKMTheorem}
{
This appendix supplies the two elementary information-geometric steps used in the main text. The microscopic label $a$ is held fixed throughout, and we work in a regulated finite-dimensional code subspace. We assume that the leading product state
\begin{equation}
\sigma_a=\rho_R^{(0)}\otimes\rho_{\mathfrak I,a}^{(0)}
\end{equation}
is faithful on its support. If zero modes are present, the argument applies after restricting to the support, or after a faithful regularization followed by the regulator limit.

\subsection*{The straight-line path and Eq.~\eqref{Eq:C-Xconn}}

The identity \eqref{Eq:C-Xconn} uses the special choice of interpolation made in \eqref{Eq:BKM-family}. Since
\begin{equation}
\rho_a(t)=\sigma_a+tX_a,
\end{equation}
its marginals are also exactly linear,
\begin{align}
\rho_{R,a}(t)&=\rho_R^{(0)}+tX_{R,a},\nonumber\\[-1mm]
\rho_{\mathfrak I,a}(t)&=\rho_{\mathfrak I,a}^{(0)}+tX_{\mathfrak I,a}.
\end{align}
Their product therefore contains a quadratic term even though the joint-state path itself is a straight line:
\begin{align}
\tau_a(t)&:=\rho_{R,a}(t)\otimes\rho_{\mathfrak I,a}(t)\nonumber\\
&=\sigma_a+tX_{a,{\rm prod}}
+t^2X_{R,a}\otimes X_{\mathfrak I,a}.
\label{Eq:AppA-product-path}
\end{align}
Subtracting this from $\rho_a(t)$ and using $X_{a,{\rm conn}}=X_a-X_{a,{\rm prod}}$ gives
\begin{equation}
\rho_a(t)-\tau_a(t)
=tX_{a,{\rm conn}}-t^2X_{R,a}\otimes X_{\mathfrak I,a}.
\label{Eq:AppA-connected-path}
\end{equation}
At the physical endpoint $t=1$, $\rho_a(1)=\rho_a^{R\mathfrak I}$ and $\tau_a(1)=\rho_{R,a}\otimes\rho_{\mathfrak I,a}$. Hence \eqref{Eq:AppA-connected-path} becomes
\begin{equation}
C_a=X_{a,{\rm conn}}-X_{R,a}\otimes X_{\mathfrak I,a},
\end{equation}
which is precisely Eq.~\eqref{Eq:C-Xconn}. In particular, $X_{a,{\rm conn}}$ is the connected correlation to linear order about the AL-saturating product state.

\subsection*{The BKM theorem: Stationarity and the BKM Hessian}

We now prove the local theorem used in \eqref{Eq:BKM-stationarity}--\eqref{Eq:BKM-Hessian}. The proof is standard; see, for example, \cite{Petz1996Monotone,LesniewskiRuskai1999}. It is useful to keep both arguments of the relative entropy explicit. Let $\rho(t)$ and $\tau(t)$ be two twice differentiable curves of density matrices meeting at the same faithful state $\sigma$ at $t=0$,
\begin{align}
\rho(t)&=\sigma+tX+\frac{t^2}{2}A+O(t^3),\nonumber\\[-1mm]
\tau(t)&=\sigma+tY+\frac{t^2}{2}B+O(t^3).
\label{Eq:AppA-general-curves}
\end{align}
Umegaki relative entropy is the Bregman divergence generated, on the affine space of unit-trace positive matrices, by
\begin{equation}
F(\rho):=\operatorname{Tr}(\rho\log\rho),
\end{equation}
namely
\begin{equation}
D(\rho\|\tau)
=F(\rho)-F(\tau)-dF_{\tau}(\rho-\tau).
\label{Eq:AppA-Bregman}
\end{equation}
For a faithful $\sigma$,
\begin{align}
dF_{\sigma}(Z)&=\operatorname{Tr}\!\left[(\log\sigma+\mathbf1)Z\right],\nonumber\\[-1mm]
d^2F_{\sigma}(Z,W)&=
\operatorname{Tr}\!\left[Z\,\Omega_{\sigma}^{-1}(W)\right],
\label{Eq:AppA-HessianF}
\end{align}
where
\begin{equation}
\Omega_{\sigma}^{-1}(W)
=\int_0^\infty ds\,(\sigma+s)^{-1}W(\sigma+s)^{-1}.
\end{equation}
The second line follows from the Fr\'echet derivative
\begin{equation}
d\log_{\sigma}(W)
=\int_0^\infty ds\,(\sigma+s)^{-1}W(\sigma+s)^{-1}.
\label{Eq:AppA-logderivative}
\end{equation}

Expanding the three terms in \eqref{Eq:AppA-Bregman} using \eqref{Eq:AppA-general-curves}, the constant and linear terms cancel. At quadratic order the terms containing the accelerations $A$ and $B$ cancel as well, leaving only the difference of the two tangents:
\begin{align}
D(\rho(t)\|\tau(t))
&=\frac{t^2}{2}\,d^2F_{\sigma}(X-Y,X-Y)+O(t^3)\nonumber\\[-1mm]
&=\frac{t^2}{2}\operatorname{Tr}\!\left[
(X-Y)\Omega_{\sigma}^{-1}(X-Y)\right]\nonumber\\[-1mm]
&\quad+O(t^3).
\label{Eq:AppA-BregmanExpansion}
\end{align}
For completeness, the cancellation can be seen directly from
\begin{align}
F(\rho(t))
&=F(\sigma)+t\,dF_{\sigma}(X)\nonumber\\[-1mm]
&\quad+\frac{t^2}{2}\bigl[dF_{\sigma}(A)+d^2F_{\sigma}(X,X)\bigr]
+O(t^3),\nonumber\\[1mm]
F(\tau(t))
&=F(\sigma)+t\,dF_{\sigma}(Y)\nonumber\\[-1mm]
&\quad+\frac{t^2}{2}\bigl[dF_{\sigma}(B)+d^2F_{\sigma}(Y,Y)\bigr]
+O(t^3).
\end{align}
The remaining Bregman term is
\begin{align}
dF_{\tau(t)}(\rho(t)-\tau(t))
&=t\,dF_{\sigma}(X-Y)\nonumber\\[-1mm]
&\quad+\frac{t^2}{2}dF_{\sigma}(A-B)\nonumber\\[-1mm]
&\quad+t^2d^2F_{\sigma}(Y,X-Y)+O(t^3).
\end{align}
which combine into \eqref{Eq:AppA-BregmanExpansion} by symmetry of the Hessian.

For the mutual information in the main text we take
\begin{equation}
\rho(t)=\rho_a(t),
\qquad
\tau(t)=\rho_{R,a}(t)\otimes\rho_{\mathfrak I,a}(t).
\end{equation}
At $t=0$ both equal $\sigma_a$, while
\begin{equation}
X-Y=X_a-X_{a,{\rm prod}}=X_{a,{\rm conn}}.
\end{equation}
Since $D(\rho_a(t)\|\tau_a(t))=\mathcal I_a(t)$, Eq.~\eqref{Eq:AppA-BregmanExpansion} yields
\begin{equation}
\dot{\mathcal I}_a(0)=0,
\end{equation}
and
\begin{equation}
\ddot{\mathcal I}_a(0)
=\operatorname{Tr}\!\left[
X_{a,{\rm conn}}\Omega_{\sigma_a}^{-1}(X_{a,{\rm conn}})\right],
\end{equation}
which are \eqref{Eq:BKM-stationarity} and \eqref{Eq:BKM-Hessian}. Positivity follows immediately in an eigenbasis $\sigma_a=\sum_i p_i|i\rangle\langle i|$ with $p_i>0$:
\begin{equation}
\operatorname{Tr}\!\left[Y\Omega_{\sigma_a}^{-1}(Y)\right]
=\sum_{i,j}|Y_{ij}|^2\,
\frac{\log p_i-\log p_j}{p_i-p_j}\geq0,
\label{Eq:AppA-positivity}
\end{equation}
with the coincident-eigenvalue limit understood as $1/p_i$.

The theorem is local: it determines the quadratic response of mutual information at $t=0$ and does not by itself justify replacing the physical value $\mathcal I_a(1)$ by that quadratic term. In the main text generalized ETH is used to control the entropy scaling of the Taylor coefficients, while the passage to $t=1$ assumes a regular order-one resummation.
}

\section{Geometric comparison of the ultraviolet and correlation scales}\label{Sec:Supp}
\addtext{This appendix gives the geometric comparison used in the main text. The two scales $L_{\rm UV}$ and $L_{\rm QC}$ are defined by different leading-HRT transitions: $L_{\rm UV}$ is fixed by the Araki--Lieb transition discussed in the main text, while $L_{\rm QC}$ measures the largest separation over which two disjoint boundary intervals can still have non-vanishing leading-HRT mutual information. We first define $L_{\rm QC}$ precisely and then show that, for a non-extremal rotating BTZ black hole, the two scales approach the same finite value as $L\to\infty$ at fixed $\mu_\pm$. Their difference is controlled by the least suppressed chiral finite-size correction and therefore behaves as $e^{-2\pi\mu_<L}=e^{-6S_</c}$.}

\subsection*{Definition of $L_{\rm QC}$ and its transition equation}
Consider two disjoint boundary intervals $R_1$ and $R_2$ separated by non-vanishing gaps of lengths $\mathbf{L}_{g_1}$ and $\mathbf{L}_{g_2}$. Their lengths satisfy
\begin{equation}
\mathbf{L}_{g_1}+\mathbf{L}_{g_2}+\mathbf{L}(R_1)+\mathbf{L}(R_2)=2\pi L
\end{equation}
(see Fig.~\ref{fig:Disjoint}). We define the leading-HRT correlation scale rigorously as
\begin{align}
L_{\rm QC}:={\rm max}\bigl\{\,&\min(\mathbf{L}_{g_1},\mathbf{L}_{g_2})\ \big|\ \nonumber\\[-1mm]
&\mathbf{L}_{g_1}+\mathbf{L}_{g_2}+\mathbf{L}(R_1)+\mathbf{L}(R_2)=2\pi L,\nonumber\\[-1mm]
&I(R_1:R_2)>0\bigr\},
\label{Eq:LQC-def}
\end{align}
where $I(R_1:R_2)$ is the mutual information computed from the leading HRRT entropies. At finite $c$, bulk mutual information can keep the exact mutual information non-zero even after the leading-HRT transition, so $L_{\rm QC}$ should not be interpreted as a sharp threshold for the exact quantum mutual information.

Using the BTZ HRRT entropies, the maximum in \eqref{Eq:LQC-def} is attained for equal intervals placed symmetrically, so that $\mathbf{L}_{g_1}=\mathbf{L}_{g_2}$. Writing
\begin{equation}
L_{\rm QC}=\pi\alpha_{\rm QC}L,
\end{equation}
the connected and disconnected HRRT saddles exchange dominance when
\begin{align}\label{Eq:LQCEq}
&2\ln\sinh(\pi\mu_+L\alpha_{\rm QC})
 +2\ln\sinh(\pi\mu_-L\alpha_{\rm QC})\nonumber\\
&\quad+2\pi L(\mu_++\mu_-)\nonumber\\
&=2\ln\sinh\!\left(\pi\mu_+L(1-\alpha_{\rm QC})\right)\nonumber\\
&\quad+2\ln\sinh\!\left(\pi\mu_-L(1-\alpha_{\rm QC})\right).
\end{align}
This is the analogue, for two disjoint intervals, of the equation \eqref{Eq:LUVEq} that determines $L_{\rm UV}$. Both equations follow from the same BTZ geodesic lengths obtained by uniformization \cite{HRT,Kibe:2024icu}. We find $L_{\rm UV}>L_{\rm QC}$, with equality reached in the large-volume limit described below.

\subsection*{Common large-volume limit and the finite-size mismatch}
It is useful to rewrite both transition equations directly in terms of the physical lengths $L_{\rm UV}$ and $L_{\rm QC}$. Using
\begin{equation}
L_{\rm UV}=2\pi\alpha_{\rm UV}L,
\qquad
L_{\rm QC}=\pi\alpha_{\rm QC}L,
\end{equation}
and
\begin{equation}
\ln\sinh z=z-\ln2+\ln(1-e^{-2z}),
\end{equation}
we define the common infinite-volume transition function
\begin{align}
\mathcal F(x)&:=\ln[\sinh(\mu_+x)\sinh(\mu_-x)]\nonumber\\[-1mm]
&\quad +(\mu_++\mu_-)x+2\ln2.
\label{Eq:F-general}
\end{align}
Then \eqref{Eq:LUVEq} and \eqref{Eq:LQCEq} are equivalently
\begin{align}
\mathcal F(L_{\rm UV})
&=\sum_{s=\pm}\ln\!\left(1-e^{-4\pi\mu_sL+2\mu_sL_{\rm UV}}\right),
\label{Eq:F-LUV}\\
\mathcal F(L_{\rm QC})
&=\sum_{s=\pm}\ln\!\left(1-e^{-2\pi\mu_sL+2\mu_sL_{\rm QC}}\right).
\label{Eq:F-LQC}
\end{align}
These formulae display the origin of the difference between the two scales. The finite-circle correction to the $L_{\rm UV}$ transition starts at $e^{-4\pi\mu_sL}$, whereas that to the $L_{\rm QC}$ transition starts already at $e^{-2\pi\mu_sL}$.

For a non-extremal black hole, $\mu_\pm>0$, both right-hand sides vanish as $L\to\infty$ at fixed $\mu_\pm$. Hence both scales approach the same finite value $L_0$ determined by
\begin{equation}
\mathcal F(L_0)=0,
\end{equation}
or, equivalently,
\begin{equation}
\bigl(e^{2\mu_+L_0}-1\bigr)\bigl(e^{2\mu_-L_0}-1\bigr)=1.
\label{Eq:L0-general}
\end{equation}
Since
\begin{align}
\mathcal F'(x)
&=\mu_+\coth(\mu_+x)+\mu_-\coth(\mu_-x)\nonumber\\[-1mm]
&\quad+\mu_++\mu_->0.
\end{align}
this positive root is unique.

We now expand about $L_0$. For fixed $0<\mu_<<\mu_>$, where
\begin{equation}
\mu_<:=\min(\mu_+,\mu_-),\qquad
\mu_>:=\max(\mu_+,\mu_-),
\end{equation}
Eqs.~\eqref{Eq:F-LUV} and \eqref{Eq:F-LQC}, together with $\ln(1-q)=-q+O(q^2)$, give
\begin{align}
L_{\rm UV}-L_0
&=-\frac{1}{\mathcal F'(L_0)}
\sum_{s=\pm}e^{2\mu_sL_0}e^{-4\pi\mu_sL}\nonumber\\[-1mm]
&\quad+o\!\left(e^{-4\pi\mu_<L}\right),
\label{Eq:LUV-expansion}\\
L_{\rm QC}-L_0
&=-\frac{1}{\mathcal F'(L_0)}
\sum_{s=\pm}e^{2\mu_sL_0}e^{-2\pi\mu_sL}\nonumber\\[-1mm]
&\quad+o\!\left(e^{-2\pi\mu_<L}\right).
\label{Eq:LQC-expansion}
\end{align}
The $L_{\rm QC}$ correction is therefore parametrically larger, and its least suppressed term comes from the smaller chiral scale $\mu_<$. Subtracting the two equations yields
\begin{equation}
L_{\rm UV}-L_{\rm QC}
=\frac{e^{2\mu_<L_0}}{\mathcal F'(L_0)}
e^{-2\pi\mu_<L}\,[1+\varepsilon(L)],
\label{Eq:general-Ldifference}
\end{equation}
where $\varepsilon(L)\to0$ as $L\to\infty$ at fixed non-zero $\mu_\pm$. Note that taking $S_<$ large is the same as taking $L$ large as $\mu_\pm$ are fixed. Thus, defining as in the main text
\begin{equation}
r:=\frac{\mu_<}{\mu_>}\in(0,1),
\end{equation}
we obtain
\begin{align}
\frac{L_{\rm UV}-L_{\rm QC}}{L_{\rm UV}}
&=A(r)e^{-2\pi\mu_<L}\,[1+\varepsilon(L)]\nonumber\\[-1mm]
&=A(r)e^{-6S_</c}\,[1+\varepsilon(L)],
\label{Eq:general-ratio}
\end{align}
where we used $S_<=(c/3)\pi L\mu_<$ and
\begin{equation}
A(r)=\frac{e^{2\mu_<L_0}}{L_0\mathcal F'(L_0)}.
\end{equation}
This is the result quoted in the main text. The coefficient $A(r)$ is independent of the circle size $L$. For example, when $r=1/2$,
\begin{equation}
L_0=\frac{1}{\mu_>}\ln\!\left(\frac{1+\sqrt5}{2}\right),
\qquad
A(1/2)\simeq0.574.
\end{equation}

At zero rotation, $\mu_+=\mu_-=\mu$, the two chiral corrections contribute at the same order. In this case
\begin{align}
L_0&=\frac{\ln2}{2\mu},\nonumber\\[-1mm]
L_{\rm UV}-L_{\rm QC}
&=\frac{e^{-2\pi\mu L}}{2\mu}
\left[1+O(e^{-2\pi\mu L})\right].
\label{Eq:appendix-q-relation}
\end{align}
so that $A(1)=1/\ln2$, in agreement with the main text.

The non-extremal expansion above is not uniform in the extremal limit $\mu_<\to0$: the factor $e^{-2\pi\mu_<L}$ ceases to be small. Correspondingly, at extremality the two scales no longer approach one another exponentially. Solving the extremal transition equations gives
\begin{equation}
L_{\rm UV}-L_{\rm QC}=\tilde{k}\,\tilde{\mu}^{-1},
\qquad \tilde{k}\approx0.33,
\end{equation}
where $\tilde\mu$ is the non-vanishing member of $\{\mu_-,\mu_+\}$. This is the geometric behavior used in the discussion of the extremal case in the main text.

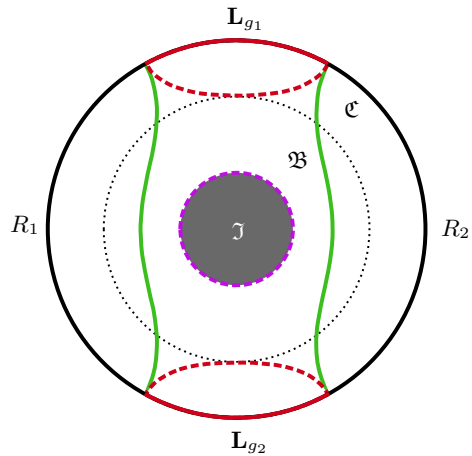
\begin{figure}
\centering   

 
\tikzset{
pattern size/.store in=\mcSize, 
pattern size = 5pt,
pattern thickness/.store in=\mcThickness, 
pattern thickness = 0.3pt,
pattern radius/.store in=\mcRadius, 
pattern radius = 1pt}
\makeatletter
\pgfutil@ifundefined{pgf@pattern@name@_zh7kgmhb3}{
\makeatletter
\pgfdeclarepatternformonly[\mcRadius,\mcThickness,\mcSize]{_zh7kgmhb3}
{\pgfpoint{-0.5*\mcSize}{-0.5*\mcSize}}
{\pgfpoint{0.5*\mcSize}{0.5*\mcSize}}
{\pgfpoint{\mcSize}{\mcSize}}
{
\pgfsetcolor{\tikz@pattern@color}
\pgfsetlinewidth{\mcThickness}
\pgfpathcircle\pgfpointorigin{\mcRadius}
\pgfusepath{stroke}
}}
\makeatother
\tikzset{every picture/.style={line width=0.75pt}} 

\begin{tikzpicture}[x=0.75pt,y=0.75pt,yscale=-1,xscale=1]
\begin{scope}[scale=0.55]
\draw  [dash pattern={on 0.75pt off 1.5pt}] (116.17,193.59) .. controls (116.17,131.92) and (166.17,81.92) .. (227.84,81.92) .. controls (289.51,81.92) and (339.5,131.92) .. (339.5,193.59) .. controls (339.5,255.26) and (289.51,305.25) .. (227.84,305.25) .. controls (166.17,305.25) and (116.17,255.26) .. (116.17,193.59) -- cycle ;
\draw  [draw opacity=0][pattern=_zh7kgmhb3,pattern size=3pt,pattern thickness=0.75pt,pattern radius=2.325pt, pattern color={rgb, 255:red, 167; green, 155; blue, 7}][line width=1.5]  (180.17,193.4) .. controls (180.17,167.18) and (201.43,145.92) .. (227.65,145.92) .. controls (253.87,145.92) and (275.13,167.18) .. (275.13,193.4) .. controls (275.13,219.62) and (253.87,240.88) .. (227.65,240.88) .. controls (201.43,240.88) and (180.17,219.62) .. (180.17,193.4)(68.91,193.4) .. controls (68.91,105.73) and (139.98,34.66) .. (227.65,34.66) .. controls (315.32,34.66) and (386.39,105.73) .. (386.39,193.4) .. controls (386.39,281.07) and (315.32,352.14) .. (227.65,352.14) .. controls (139.98,352.14) and (68.91,281.07) .. (68.91,193.4) ;
\draw [color={rgb, 255:red, 60; green, 190; blue, 31 }  ,draw opacity=1 ][line width=1.5]    (151.61,332.03) .. controls (154.86,324.28) and (156.61,320.14) .. (158.11,314.14) .. controls (159.61,308.14) and (159.89,305.03) .. (160.36,299.89) .. controls (160.82,294.75) and (160.86,288.64) .. (160.86,285.91) .. controls (160.86,283.18) and (159.11,265.18) .. (157.11,257.68) .. controls (155.11,250.18) and (152.61,237.68) .. (151.36,231.43) .. controls (150.11,225.18) and (148.61,216.68) .. (148.11,211.96) .. controls (147.61,207.23) and (146.86,194.71) .. (147.11,190.23) .. controls (147.36,185.75) and (148.11,175.25) .. (148.86,170.03) .. controls (149.61,164.8) and (152.54,148.67) .. (155.11,138.03) .. controls (157.68,127.38) and (158.13,124.22) .. (159.36,115.57) .. controls (160.59,106.92) and (160.86,99.82) .. (160.86,93.32) .. controls (160.86,86.82) and (159.36,77.82) .. (157.61,70.82) .. controls (155.86,63.82) and (153.86,60.07) .. (151.36,55.37) ;
\draw [color={rgb, 255:red, 60; green, 190; blue, 31 }  ,draw opacity=1 ][line width=1.5]    (304.29,331.75) .. controls (301.28,325.36) and (298.35,317.56) .. (296.79,310.5) .. controls (295.23,303.45) and (295.04,300.16) .. (294.79,292.3) .. controls (294.55,284.44) and (295.2,277.76) .. (296.29,271.03) .. controls (297.39,264.29) and (299.04,256.28) .. (300.04,251.28) .. controls (301.04,246.28) and (303.68,235.4) .. (304.79,229.05) .. controls (305.9,222.7) and (306.79,217.87) .. (307.54,211.12) .. controls (308.29,204.37) and (308.48,201.69) .. (308.54,195.64) .. controls (308.6,189.59) and (308.16,183.81) .. (308.04,181.12) .. controls (307.92,178.43) and (306.42,166.2) .. (304.79,157.66) .. controls (303.17,149.13) and (301.4,141.46) .. (300.29,136.66) .. controls (299.18,131.86) and (297.94,126.2) .. (296.79,118.96) .. controls (295.64,111.71) and (295.28,106.47) .. (295.04,101.21) .. controls (294.8,95.94) and (295.08,92.43) .. (295.54,86.46) .. controls (296,80.48) and (296.6,75.68) .. (297.79,71.98) .. controls (298.98,68.28) and (301.63,59.84) .. (304.29,55.21) ;
\draw  [color={rgb, 255:red, 208; green, 2; blue, 27 }  ,draw opacity=1 ][fill={rgb, 255:red, 255; green, 255; blue, 255 }  ,fill opacity=1 ][dash pattern={on 3pt off 1.5pt}][line width=1.5]  (179.65,311.53) .. controls (184.15,310.03) and (195.15,307.28) .. (207.4,306.53) .. controls (219.65,305.78) and (225.05,305.83) .. (234.65,305.78) .. controls (244.26,305.72) and (250.4,306.78) .. (256.15,307.28) .. controls (261.9,307.78) and (266.43,308.48) .. (273.7,310.64) .. controls (280.97,312.8) and (285.7,314.64) .. (290.7,317.89) .. controls (295.7,321.14) and (297.95,323.4) .. (300.7,326.64) .. controls (303.45,329.87) and (304.23,331.97) .. (304.52,332.16) .. controls (304.8,332.35) and (301.45,334.49) .. (293.95,337.74) .. controls (286.45,340.99) and (284.45,341.99) .. (278.2,343.99) .. controls (271.95,345.99) and (264.7,347.99) .. (256.95,349.49) .. controls (249.2,350.99) and (239.45,351.99) .. (227.84,352.14) .. controls (216.23,352.29) and (205.09,350.84) .. (195.52,348.91) .. controls (185.95,346.99) and (176.95,343.99) .. (169.45,340.99) .. controls (161.95,337.99) and (150.77,333.06) .. (151.61,332.03) .. controls (152.45,330.99) and (154.98,325.99) .. (160.65,321.53) .. controls (166.33,317.06) and (175.15,313.03) .. (179.65,311.53) -- cycle ;
\draw  [color={rgb, 255:red, 208; green, 2; blue, 27 }  ,draw opacity=1 ][fill={rgb, 255:red, 255; green, 255; blue, 255 }  ,fill opacity=1 ][dash pattern={on 3pt off 1.5pt}][line width=1.5]  (152.02,54) .. controls (152.21,54.12) and (156.81,51.25) .. (161.08,49.39) .. controls (165.34,47.52) and (171.29,45.16) .. (174.52,43.75) .. controls (177.75,42.35) and (193.64,38.16) .. (196.58,37.64) .. controls (199.51,37.11) and (206.78,35.92) .. (211.95,35.39) .. controls (217.12,34.86) and (222.09,34.78) .. (228.09,34.66) .. controls (234.09,34.53) and (253.58,36.39) .. (257.58,37.39) .. controls (261.58,38.39) and (272.83,40.64) .. (283.58,44.64) .. controls (294.33,48.64) and (305.83,54.64) .. (304.88,54.7) .. controls (303.93,54.77) and (303.91,56.05) .. (301.77,59.25) .. controls (299.63,62.45) and (297.63,63.7) .. (295.88,65.2) .. controls (294.13,66.7) and (289.63,69.45) .. (286.27,71.25) .. controls (282.91,73.05) and (281.63,73.7) .. (278.2,74.89) .. controls (274.77,76.07) and (269.63,77.57) .. (262.02,78.75) .. controls (254.41,79.93) and (244.28,80.75) .. (235.7,81.14) .. controls (227.13,81.52) and (207.52,80.5) .. (198.77,79.5) .. controls (190.02,78.5) and (178.63,74.78) .. (174.45,73.14) .. controls (170.27,71.5) and (167.27,69.75) .. (163.27,67) .. controls (159.27,64.25) and (156.77,62) .. (155.52,60.25) .. controls (154.27,58.5) and (151.83,53.89) .. (152.02,54) -- cycle ;
\draw  [color={rgb, 255:red, 0; green, 0; blue, 0 }  ,draw opacity=1 ][line width=1.5]  (69.1,193.4) .. controls (69.1,105.73) and (140.17,34.66) .. (227.84,34.66) .. controls (315.51,34.66) and (386.58,105.73) .. (386.58,193.4) .. controls (386.58,281.07) and (315.51,352.14) .. (227.84,352.14) .. controls (140.17,352.14) and (69.1,281.07) .. (69.1,193.4) -- cycle ;
\draw  [color={rgb, 255:red, 189; green, 16; blue, 224 }  ,draw opacity=1 ][fill={rgb, 255:red, 0; green, 0; blue, 0 }  ,fill opacity=0.59 ][dash pattern={on 3pt off 1.5pt}][line width=1.5]  (180.55,193.4) .. controls (180.55,167.28) and (201.72,146.11) .. (227.84,146.11) .. controls (253.95,146.11) and (275.13,167.28) .. (275.13,193.4) .. controls (275.13,219.51) and (253.95,240.68) .. (227.84,240.68) .. controls (201.72,240.68) and (180.55,219.51) .. (180.55,193.4) -- cycle ;
\draw [color={rgb, 255:red, 208; green, 2; blue, 27 }  ,draw opacity=1 ][line width=1.5]    (150.58,332.14) .. controls (158.83,336.39) and (161.08,337.89) .. (175.58,343.14) .. controls (190.08,348.39) and (208.95,352.14) .. (224.2,352.14) .. controls (239.45,352.14) and (254.7,351.14) .. (272.7,345.64) .. controls (290.7,340.14) and (292.2,338.39) .. (305.7,331.89) ;
\draw [color={rgb, 255:red, 208; green, 2; blue, 27 }  ,draw opacity=1 ][line width=1.5]    (151.08,54.39) .. controls (153.8,52.89) and (161.66,48.99) .. (166.33,47.14) .. controls (170.99,45.29) and (177.33,42.89) .. (184.83,40.64) .. controls (192.33,38.39) and (196.08,37.64) .. (203.83,36.39) .. controls (211.58,35.14) and (218.83,34.39) .. (227.33,34.64) .. controls (235.83,34.89) and (240.85,34.89) .. (249.83,36.14) .. controls (258.8,37.39) and (267.83,39.39) .. (274.58,41.64) .. controls (281.33,43.89) and (284.14,44.82) .. (290.33,47.39) .. controls (296.51,49.96) and (301.93,52.67) .. (304.58,54.39) ;

\draw (221.41,185.68) node [anchor=north west][inner sep=0.75pt]  [color={rgb, 255:red, 255; green, 255; blue, 255 }  ,opacity=1 ]  {$\mathfrak{I}$};
\draw (35,179.6) node [anchor=north west][inner sep=0.75pt]    {$R_{1}$};
\draw (397.2,182.8) node [anchor=north west][inner sep=0.75pt]    {$R_{2}$};
\draw (268.23,125.73) node [anchor=north west][inner sep=0.75pt]    {$\mathfrak{B}$};
\draw (315.73,83.23) node [anchor=north west][inner sep=0.75pt]    {$\mathfrak{C}$};
\draw (217.2,5) node [anchor=north west][inner sep=0.75pt]    {$\mathbf{L}_{g_{1}}$};
\draw (220.4,361) node [anchor=north west][inner sep=0.75pt]    {$\mathbf{L}_{g_{2}}$};

\end{scope}
\end{tikzpicture}

    \caption{Two non-overlapping intervals $R_1$ and $R_2$ at the boundary are shown above with two non-vanishing gaps $ \mathbf{L}_{g1}$ and $ \mathbf{L}_{g2}$ between them. The mutual information between $R_1$ and $R_2$ is non-vanishing if the bulk entanglement wedge is connected (the dotted region) bounded by the dotted red bulk geodesics and the outer horizon instead of the union of the two disconnected regions bounded by the green bulk geodesics. Here the BTZ black hole shown with compactified radial coordinate $r\to \cot\rho$ has $\mu_+=\mu_-=0.25$ (AdS radius $\ell =1$).}
    \label{fig:Disjoint}
\end{figure}

\section{Explicit examples and the limit $L\rightarrow\infty$}\label{Sec:Ex}
For $\mu_+= \mu_- =\mu$ (no rotation), 
\begin{align*}
    &L_{\rm UV} =\frac{{\rm arccoth}[1+ 2\coth(2\pi\mu L)]}{\mu },\nonumber\\
    &L_{\rm QC} =\frac{{\rm arccoth}[1+ 2\coth(\pi\mu L)]}{\mu }.
\end{align*}
If we keep $\mu$ fixed and take the length of the boundary circle to infinity i.e. $L \rightarrow\infty$ (or equivalently $S_{BH}\rightarrow \infty$ at fixed $\beta$), then in this case
\[
L_{\rm UV} = L_{\rm QC} = \frac{\ln 2}{2\mu}\sim \frac{0.35}{\mu}.
\]
For zero rotation $L_{\rm IR}=\sqrt{\beta_+\beta_-}=\beta=\pi/\mu$, and therefore in the above limit $L_{\rm UV}/L_{\rm IR} = L_{\rm QC}/L_{\rm IR} = \ln 2/2\pi \sim 0.11$. Note that the entanglement wedge transition (when the HRRT surface transits to the disconnected saddle) for a boundary interval of length $l$ occurs when $l > 2\pi L - L_{\rm UV}$ which is infinite although $L_{\rm UV}$ and $L_{\rm QC}$ are both remarkably finite in this limit.  Note that the corona is well defined in this limit as $L_{\rm UV}$ is finite. (Clearly in the $\mu\rightarrow 0$ limit at finite $L$ (global AdS$_3$), we get $L_{\rm UV} = L\pi$ and $L_{\rm QC} = L\pi/2$.)

\textit{Generally, the corona is well defined even in the limit in which $L\rightarrow\infty$ at fixed $\mu_+$ and $\mu_-$ ($S_{BH}\rightarrow\infty$ at fixed temperatures of the left/right movers)} as in the non-rotating BTZ black hole case discussed above. As for instance, when $\mu_</\mu_> = 1/2$, we obtain that
\[
L_{\rm UV} = L_{\rm QC} = \frac{\ln\left(\frac{1+\sqrt{5}}{2}\right)}{\mu_>}\sim \frac{0.48}{\mu_>}
\]
while \addtext{$L_{\rm IR}=\sqrt{\beta_+\beta_-}=\sqrt{2}\,\pi/\mu_>$, so $L_{\rm UV}/L_{\rm IR}=\ln[(1+\sqrt5)/2]/(\sqrt2\pi)\simeq0.108$} in the limit $L\rightarrow \infty$ at fixed $\mu_>$.

\bibliographystyle{elsarticle-num}
\bibliography{References}

\end{document}